\documentclass[pra, twocolumn,showpacs,nofootinbibfloatfix,amsmath,amsfonts,amssymb]{revtex4-1}%
\usepackage{amsmath,amsfonts,amssymb,color}
\usepackage{amsthm}
\usepackage{leftidx}
\usepackage{graphicx}
\usepackage{xcolor}
\usepackage{dcolumn}
\usepackage{bm}
\usepackage{epstopdf}
\usepackage{epsfig}
\usepackage{mathdots} 
\def\RB{\textcolor{black}}

\begin{document}
	\title{Generalized Majorana edge modes in a number-conserving periodically driven $p$-wave superconductor}
	\author{Raditya Weda Bomantara}
	\email{Raditya.Bomantara@kfupm.edu.sa}
	\affiliation{%
		Department of Physics, Interdisciplinary Research Center for Intelligent Secure Systems, King Fahd University of Petroleum and Minerals, 31261 Dhahran, Saudi Arabia
	}
	\date{\today}
	
	%%%%%%%%%%%%%%%%%%%% ABSTRACT %%%%%%%%%%%%%%%%%%%%%%%%
	%\begin{linenumbers}
	
	\vspace{2cm}
	
\begin{abstract}
We study an analytically solvable and experimentally relevant number-conserving periodically driven $p$-wave superconductor. Such a system is found to support generalized Majorana zero and $\pi$ modes which, despite being non-Hermitian, are still capable of encoding qubits. Moreover, appropriate winding numbers characterizing the topology of such generalized Majorana modes are defined and explicitly calculated. We further discuss the fate of the obtained generalized Majorana modes in the presence of finite charging energy. Finally, we shed light on the quantum computing prospects of such modes by demonstrating the robustness of their encoded qubits and explicitly braiding a pair of generalized Majorana modes.
	
\end{abstract}

\maketitle

\section{Introduction} 
\label{intro}

Since the seminal work of Kitaev \cite{Kit}, $p$-wave superconductors have been extensively studied for their potential ability to host the sought-after Majorana zero energy excitations at their edges. Such Majorana zero modes (MZMs), which are of interest to condensed matter and quantum computing communities alike, are exotic quasiparticles that could be understood as the building blocks of fermions \cite{Aguado2017}. In particular, two spatially separated MZMs at the edges of a $p$-wave superconductor form a nonlocal fermion that could potentially be utilized as a highly resilient qubit. Such a qubit could in turn be manipulated by moving its MZM constituents around one another, a process referred to as braiding \cite{Ivanov2000, Nayak2008}. Remarkably, quantum gate operations that result from braiding processes are topologically protected and are thus expected to be robust against errors and imperfections \cite{Lahtinen2017}. For this reason, Majorana-based qubits are considered one of the most promising building-blocks of fault-tolerant quantum computers.

At present, significant theoretical advances have been made towards harnessing the robust quantum computational power of MZMs. Indeed, the set of all topologically protected quantum gate operations realizable from braiding of MZMs have been identified \cite{Ahlbrecht2009}, whilst the potential realization of braiding itself has been proposed by several groups since the last decade \cite{braid1,braid2,braid3,braid4}. In recent years, measurement-only braiding protocols that do not involve physically moving MZMs were further developed \cite{mbraid1,mbraid2,mbraid3,mbraid4}. Such protocols have in turn stimulated various proposals for Majorana-qubit architecture designs \cite{march1,march2,march3} and Majorana-based quantum error corrections \cite{msur1,msur2,msur3,msur4}.  

\RB{Being the main ingredient of MZMs, true (spin-triplet) $p$-wave superconductivity is very scarce in nature. In the last decade, two independent groups proposed a means to artificially engineer $p$-wave superconductivity by proximitizing a more common spin-singlet $s$-wave superconductor to a semiconducting wire of finite spin-orbit coupling under the influence of a magnetic field \cite{exprop1,exprop2}. Such a proposal has since been implemented in the lab by various experimental groups \cite{mexp1,mexp2,mexp3,mexp4,mexp5}.} Despite these extensive efforts, however, an unambiguous signature of MZMs has yet to be observed. 

The very large gap between theoretical and experimental progresses above could be attributed to a number of factors. Among others, that many existing theoretical studies employ the mean-field descriptions of $p$-wave superconductors may lead to difficulties in establishing their exact correspondence with actual experimental systems. For this reason, several works have been devoted towards exploring the full interacting version of $p$-wave superconductors in recent years \cite{Fidkowski2011,Sau2011,Cheng2011,Kraus2013,Ortiz2014,Ortiz2016,Wang2017,Lapa2020a,Lapa2020}. As compared to their mean-field counterparts, which obey the parity ($\mathbb{Z}_2$) conservation law, such systems obey the more physical number conservation law. This radical change in the systems' global symmetry may in turn affect the properties of MZMs therein and, potentially, their ability to perform as topological qubits. Unfortunately, the necessity to account for interaction effect in such number-conserving $p$-wave superconductors makes their characterizations significantly harder as compared to their mean-field counterparts. As a result, many existing works rely on sophisticated techniques such as Density Matrix Renormalization Group \cite{Kraus2013} and bosonization \cite{Fidkowski2011,Sau2011,Cheng2011}, which however obscure any intuitive physical pictures.    

A rigorous analytical investigation of a particular number-conserving $p$-wave superconducting model was recently made in Ref.~\cite{Lapa2020}. Specifically, the model describes a one-dimensional (1D) fermionic lattice interacting with a bulk superconductor, which is thus of relevance to existing experiments involving semiconductor-superconductor heterostructures \cite{mexp1,mexp2,mexp3,mexp4,mexp5}. Main results of reference~\cite{Lapa2020} include the emergence of Majorana-like zero modes at the edges of the system which are no longer Hermitian like their mean-field counterparts, the finite energy gap between the unique ground state and the excited states at a fixed particle number sector, and the preservation of the corresponding Majorana-like parity. 

In this paper, we analytically and numerically investigate a periodically driven version of the number-conserving $p$-wave superconductor studied in Ref.~\cite{Lapa2020}. Our work is especially motivated by the knowledge that time-periodic mean-field $p$-wave superconductors may harbor the so-called Majorana $\pi$ modes (MPMs) that have no static counterparts \cite{FMF1,FMF2,FMF3,FMF6,FMF7,Sen1,Sen2}. As the name suggests, MPMs are another type of Majorana quasiparticle excitations that occurs at half the driving frequency \cite{FMFrev}. In particular, as MPMs may emerge simultaneously with the more common MZMs, time-periodic $p$-wave superconductors have the potential to support more qubits than their static counterparts under the same physical resource \cite{FMF5,RG,RG2020b,Matthies2022}. In addition, there exists a means to geometrically braid an MZM and an MPM in a strictly 1D setting \cite{FMF4,Bauer2018}, thus bypassing the necessity for engineering a quasi-1D design that may present a number of issues \cite{Nijholt2016}. 

To our knowledge, all existing studies on time-periodic $p$-wave superconductors and the corresponding formation of MPMs assume mean-field treatment. The fate and quantum computational abilities of Majorana modes in number-conserving time-periodic $p$-wave superconductors have so far remained unexplored. This paper thus aims to initiate and stimulate studies along this direction. Our main findings demonstrate that a generalization of both MZMs and MPMs emerge in the number-conserving setting. Due to the simplicity of the model under consideration, we are able to analytically derive the exact expressions for the generalized Majorana modes under a semi-infinite geometry, and subsequently derive the conditions under which they exist. In addition, we manage to establish bulk-edge correspondence by analytically defining and evaluating the appropriate topological invariants under a closed geometry. Remarkably, we further show numerically that braiding the obtained generalized Majorana modes in a strictly 1D setup is possible, thus serving as a first step towards harnessing their quantum computational power.

This paper is organized as follows. In Sec.~\ref{model}, we present the main Hamiltonian analyzed in this paper, which describes a number-conserving periodically driven $p$-wave superconductor, and highlight the conserved number operator. We then start our investigation with the time independent version of the Hamiltonian in Sec.~\ref{model}A, where we do not only verify the results of Ref.~\cite{Lapa2020} both numerically and via a different analytical method, but we also establish the topological nature of the system by explicitly defining and computing an appropriate topological invariant. The full time-periodic version of the Hamiltonian is then analyzed in Sec.~\ref{model}B, where we obtain and topologically characterize a generalization of the MPMs that may simultaneously coexist with generalized MZMs. In Sec.~\ref{finEc}, we discuss the effect of finite charging energy on the previously obtained generalized Majorana modes. In Sec.~\ref{discdis}, we demonstrate the robustness of the generalized Majorana modes against disorders, whereas the robustness of the qubits they encode at finite charging energy is the subject of Sec.~\ref{discchar}. We further develop and explicitly execute a braiding protocol between two generalized Majorana modes in Sec.~\ref{discbraid}. Finally, we conclude this manuscript and present potential future directions in Sec.~\ref{conc}.

\section{number-conserving periodically driven topological superconductor}
\label{model}

We consider a number-conserving variation of a periodically driven $p$-wave superconductor described by the Hamiltonian:
	\begin{eqnarray}
	H(t)&=&\sum_{j=1}^{L-1} \left(J(t) c_{j+1}^\dagger c_j + \Delta(t) e^{i\hat{\phi}}  c_{j} c_{j+1} +h.c. \right)  \nonumber \\
      && + \RB{\mu(t) \left(\sum_{j=1}^Lc_j^\dagger c_j-\frac{L}{2}\right)} + E_c (2\hat{n}-2n_c)^2 \;. \label{ncons}
	\end{eqnarray}
It physically represents a 1D fermionic lattice in proximity to a bulk $p$-wave superconductor, in which the degrees of freedom of both the 1D lattice and the superconductor reservoir are taken into account \cite{Lapa2020}. There, $J(t)$, $\mu(t)$, and $\Delta(t)$ are respectively the time periodic hopping, chemical potential, and $p$-wave superconducting pairing strength, $c_j$ ($c_j^\dagger$) is the fermionic annihilation (creation) operator acting on site $j$, $L$ is the system size, $\hat{n}$ describes the number operator of Cooper pairs on the reservoir, $E_c$ is the charging energy, $n_c$ is the reference number of Cooper pairs in the reservoir, and $e^{i\hat{\phi}}$ is the Cooper-pair creation operator which satisfies $[\hat{n},e^{i\hat{\phi}}] = e^{i\hat{\phi}}$. In particular, Eq.~(\ref{ncons}) commutes with the operator 
\begin{equation}
    \hat{M} = \sum_{j=1}^L c_{j}^\dagger c_j + 2\hat{n} , \label{tnum}
\end{equation}
which counts the total number of particles in the lattice and reservoir combined. In Sec.~\ref{ticase} and \ref{tpcase}, we analyze the topological aspects of Eq.~(\ref{ncons}) in the special case of $E_c=0$. The effect of finite charging energy $E_c$ will be elucidated in Sec.~\ref{finEc}.  

\subsection{Time-independent case}
\label{ticase}

A time-independent version of Eq.~(\ref{ncons}), which has been rigorously studied in Ref.~\cite{Lapa2020}, amounts to setting $J(t)=J_0$, $\Delta(t)=\Delta_0$, and $\mu(t)=\mu_0$. In particular, by considering the special case of $\mu_0=0$ and $J_0=\Delta_0$, the system's Hamiltonian commutes with the two edge operators $\hat{\Gamma}_{A,1}\equiv e^{\mathrm{i}\hat{\phi}} c_1+c_1^\dagger$ and $\hat{\Gamma}_{B,L}\equiv \mathrm{i} \left( e^{\mathrm{i}\hat{\phi}} c_L- c_L^\dagger \right)$ \cite{Lapa2020}, which could be regarded as the number-conserving generalizations of MZMs in the mean-field model. Indeed, by defining a set of mutually anti-commuting operators $\hat{\Gamma}_{A,j} \equiv e^{\mathrm{i}\hat{\phi}} c_j+c_j^\dagger$ and $\hat{\Gamma}_{B,j}\equiv \mathrm{i} \left( e^{\mathrm{i}\hat{\phi}} c_j- c_j^\dagger \right)$, Eq.~(\ref{ncons}) under these special parameters can be written in the form
\begin{equation}
     H_{\rm stat} = -\sum_{j=1}^{L-1} J_0 \mathrm{i} \hat{\Gamma}_{A,j+1}^\dagger \hat{\Gamma}_{B,j} . \label{nconsstat} 
\end{equation}
As $\hat{\Gamma}_{A,1}$ and $\hat{\Gamma}_{B,L}$ are absent in Eq.~(\ref{nconsstat}), they commute with and are thus zero modes of $H_{\rm stat}$. Reference~\cite{Lapa2020} further shows that within a fixed total number of particles, i.e., $\hat{M}=M$, the system's ground state is well-gapped from the excited states. 

In this subsection, we develop an alternative framework to verify these findings. Our framework has the advantage of clearly demonstrating the topological nature of the obtained zero modes. Moreover, the same framework could be easily adapted to investigate the presence of similar generalized Majorana edge modes in the more general time-periodic case, which will be the subject of the next subsection. We start by writing Eq.~(\ref{ncons}) in a Bogoliubov-de Gennes-(BdG-)like form as
\begin{equation}
    H = \frac{1}{2} \psi^\dagger \mathcal{H} \psi ,
\end{equation}
where $\psi= (c_1, c_2, \cdots , c_{N}, c^\dagger_1, \cdots, c^\dagger_N)^T$ is the Nambu vector. The infinite dimensional BdG matrix $\mathcal{H}$ can be written as 
\begin{equation}
    \mathcal{H} = J_0 \sigma_z \eta_x +i \Delta_0  (e^{-\mathrm{i} \hat{\phi}} \sigma_+ - e^{\mathrm{i} \hat{\phi}} \sigma_- ) \eta_y + \mu_0 \sigma_z \eta_0 ,  \label{nconsstatbdg}
\end{equation}
where $\sigma_{x/y/z}$ are the usual Pauli matrices, $\sigma_\pm = \sigma_x \pm i \sigma_y$, $\eta_0$ and $\eta_{x/y}$ are $N\times N$ matrices acting on lattice site degree of freedom which respectively act as identity, $[\eta_x]_{j,j'}= \delta_{j + 1,j'}+\delta_{j - 1,j'}$, and $[\eta_y]_{j,j'} = -i\delta_{j + 1,j'}+i\delta_{j - 1,j'}$.  

In the same spirit of the mean-field case, the energy excitation spectrum of our number-conserving model can be obtained by diagonalizing Eq.~(\ref{nconsstatbdg}). In particular, generalized MZMs (gMZMs) are obtained from the zero eigenvalue solutions of $\mathcal{H}$ that are localized near one of the system ends. As detailed in Appendix~\ref{app:A}, in the special case $J_0=\Delta_0$, gMZMs exist whenever $|\mu|<|2J_0|$ and are approximately given by 
\begin{eqnarray}
    %\hat{\Gamma}_L &\approx& \sum_{j=1}^L \left(-\frac{\mu_0}{2J_0} \right)^{j-1} \left(c_j^\dagger + e^{\mathrm{i} \hat{\phi}}c_j \right) , \nonumber  \\
    %\hat{\Gamma}_R &\approx& \sum_{j=1}^L \left(-\frac{\mu_0}{2J_0} \right)^{j-1} \mathrm{i} \left(c_{L-j+1}^\dagger - e^{\mathrm{i} \hat{\phi}}c_{L-j+1} \right) ,
    \hat{\Gamma}_L &\approx& \sum_{j=1}^L \left(-\frac{\mu_0}{2J_0} \right)^{j-1} \hat{\Gamma}_{A,j} , \nonumber  \\
    \hat{\Gamma}_R &\approx& \sum_{j=1}^L \left(-\frac{\mu_0}{2J_0} \right)^{j-1} \hat{\Gamma}_{B,L-j+1} ,
    \label{exactMM}
\end{eqnarray}
which become exact in the limit $L\rightarrow \infty$. In particular, at $\mu_0=0$, $\hat{\Gamma}_L$ and $\hat{\Gamma}_R$ respectively reduce to $\hat{\Gamma}_{A,1}$ and $\hat{\Gamma}_{B,L}$ correctly. It is worth noting that, unlike MZMs that emerge in the mean-field model, the gMZMs $\hat{\Gamma}_L$ and $\hat{\Gamma}_R$ are not Hermitian and are thus not Majorana operators. However, as discussed in Sec.~\ref{discchar}, such gMZMs could still be utilized to encode a nonlocal qubit in the same spirit of true MZMs.

%unlike MZMs that emerge in the mean-field model, the gMZMs $\hat{\Gamma}_L$ and $\hat{\Gamma}_R$ are not Hermitian and are thus not Majorana operators. However, a nonlocal generalized Majorana parity operator could still be defined as $P=\mathrm{i}\hat{\Gamma}_L^\dagger \hat{\Gamma}_R$. In particular, as $P^2=1$, $P$ has $\pm 1$ eigenvalues. As $[H,P]=[H,\hat{\Gamma}_L]=[H,\hat{\Gamma}_R]=0$ but $\left\lbrace P, \hat{\Gamma}_L\right\rbrace = 0$, eigenvalues of $H$ are guaranteed to be at least two-fold degenerate, the eigenstates of which can further be labelled by the eigenvalues of $P$ \cite{Fendley2012}. Therefore, despite not being strictly Majorana in nature the gMZMs $\hat{\Gamma}_L$ and $\hat{\Gamma}_R$ are expected to inherit the quantum computational advantages of true MZMs of mean-field topological superconductors.   

The topological nature of the above obtained gMZMs could also be uncovered by analyzing the system's symmetry and defining an appropriate topological invariant. Under periodic boundary conditions (PBC), we may further write
\begin{equation}
    H = \sum_{k} \frac{1}{2} \psi_k^\dagger \mathcal{H}_k \psi_k ,
\end{equation}
where $\psi_k= (c_k, c_{-k}^\dagger)^T$ and $\mathcal{H}_k$ is obtained from Eq.~(\ref{nconsstatbdg}) under the replacement of $\eta_x \rightarrow 2\cos(k)$ and $\eta_y \rightarrow 2\mathrm{i} \sin(k) $. It is then easily verified that the system respects the particle-hole symmetry $\mathcal{P} \mathcal{H}_k \mathcal{P}^\dagger = - \mathcal{H}_{-k} $, where $\mathcal{P}=\mathcal{K} \sigma_x$ and $\mathcal{K}$ is the complex conjugation operator. The system also respects two chiral symmetries $\mathcal{C}_1 \mathcal{H}_k \mathcal{C}_1^\dagger = - \mathcal{H}_{k}$ and $\mathcal{C}_2 \mathcal{H}_k \mathcal{C}_2^\dagger = - \mathcal{H}_{k}$ with $\mathcal{C}_1= \Phi_x \sigma_y$ and $\mathcal{C}_2= \Phi_y \sigma_x$, where $\Phi_x$ and $\Phi_y$ are operators acting on the Cooper-pair subspace that are explicitly given by (in the Cooper-pair number basis) \cite{RG2020},
\begin{eqnarray}
    \;[\Phi_x]_{ab} &=& \delta_{1-a,b} , \nonumber \\
    \;[\Phi_y]_{ab} &=& (-\mathrm{i})^{2b-1} \delta_{1-a,b} .
\end{eqnarray}
\RB{In particular, $\Phi_x$ and $\Phi_y$ are defined as the infinite dimensional generalizations of the Pauli matrices which are simultaneously Hermitian, unitary, and squaring to identity. Moreover, they anticommute with each other and satisfy the commutation relations (as verified in Appendix~\ref{app:exB})} $\Phi_x \cos \hat{\phi} \Phi_x = -\Phi_y \cos \hat{\phi} \Phi_y =\cos\hat{\phi}$ and $-\Phi_x \sin \hat{\phi} \Phi_x = \Phi_y \sin \hat{\phi} \Phi_y =\sin\hat{\phi}$. The system consequently also respects two time-reversal symmetries $\mathcal{T}_1 \mathcal{H}_k \mathcal{T}_1^\dagger =  \mathcal{H}_{-k}$ and $\mathcal{T}_2 \mathcal{H}_k \mathcal{T}_2^\dagger =  \mathcal{H}_{-k}$ with $\mathcal{T}_1=\mathcal{P}\mathcal{C}_1$ and $\mathcal{T}_2=\mathcal{P}\mathcal{C}_2$. This places the system in class BDI of Altland-Zirnbauer symmetry classification \cite{AZ}, whose topology is determined by a $\mathbb{Z}$ invariant.

\begin{center}
\begin{figure}
    \includegraphics[scale=0.42]{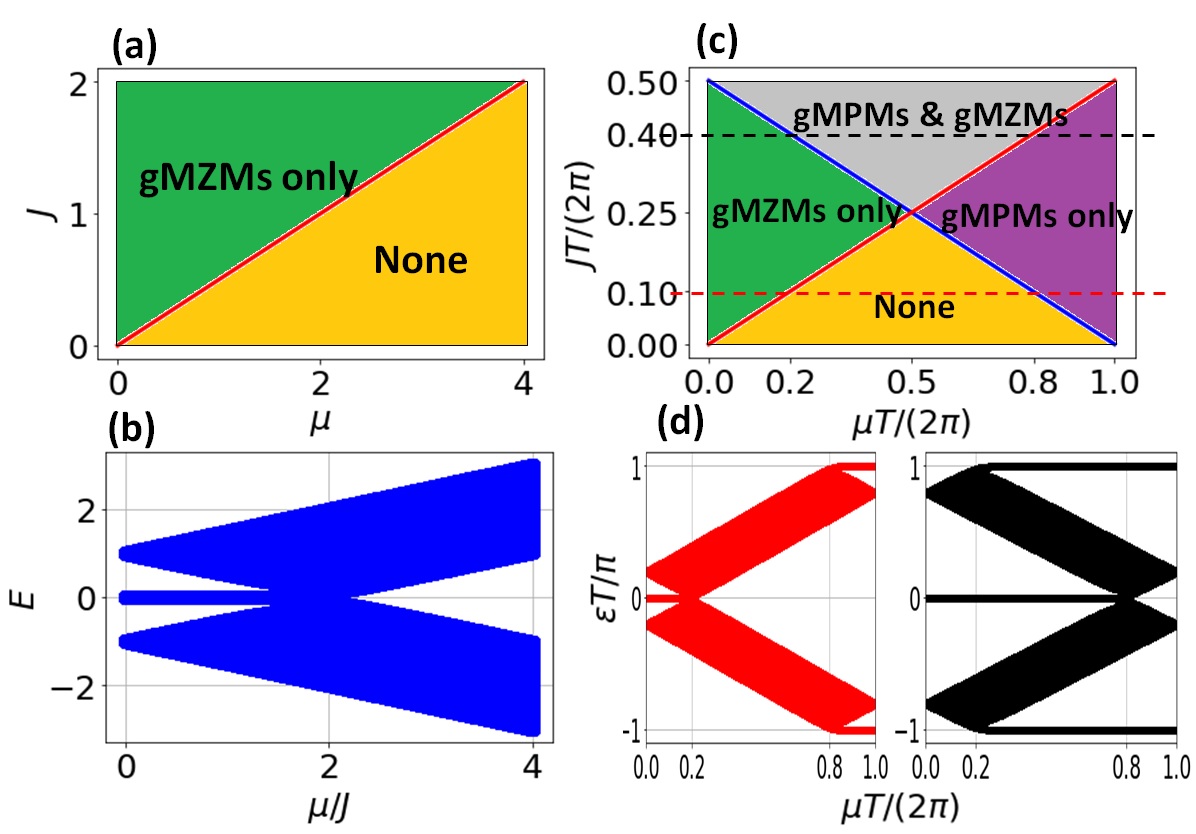}
    \caption{(a,c) The phase diagram of our number-conserving $p$-wave superconductors \RB{from calculating the appropriate normalized winding numbers} under (a) static parameters $J_0=\Delta_0= J$ and $\mu_0= \mu$, and (c) time-periodic parameters of Sec.~\ref{tpcase}. (b) The energy excitation spectrum of the static number-conserving $p$-wave superconducting system discussed in Sec.~\ref{ticase}. (d) The quasienergy excitation spectrum of the periodically driven number-conserving $p$-wave superconductor discussed in Sec.~\ref{tpcase}, at a fixed $JT=\Delta T = 0.2\pi$ (left subpanel) and $JT=\Delta T = 0.8\pi$ (right subpanel). In panel (c), the red and black horizontal lines show the parameters used in panel (d). In panels (b) and (d), the system size is taken as $L=15$, $E_c=0$, and the Cooper-pairs degree of freedom is truncated at $n_{\rm min} = -10$ and $n_{\rm max} = 10$.}
    \label{fig:mzphase}
\end{figure}    
\end{center}

\RB{A suitable $\mathbb{Z}$ invariant for our system can be constructed by first writing the matrix representation for $\mathcal{H}_k$ in the basis where $\mathcal{C}_1$ is diagonal. In this case, $\mathcal{H}_k$ also takes the anti-diagonal form
\begin{equation}
    \mathcal{H}_k \rightarrow \left( \begin{array}{cc}
        0 & \mathcal{W} \\
        \RB{\mathcal{W}^\dagger} & 0 
    \end{array} \right) ,
\end{equation}
from which we can define the normalized winding number as
\begin{equation}
    w= \frac{1}{2\pi \mathrm{i} \mathrm{Tr}(\mathcal{I}_c)} \oint \mathrm{Tr}\left(\mathcal{W}^{-1}d\mathcal{W}\right),
\end{equation}
where $\mathcal{I}_c$ is the identity in the Cooper-pairs subspace. Here, the division over $\mathrm{Tr}(\mathcal{I}_c)$ is considered to avoid the possible infinity value arising from $\mathrm{Tr}\left(\mathcal{W}^{-1}d\mathcal{W}\right)$, i.e., the trace is taken over the infinite dimensional Cooper-pairs subspace.} As detailed in Appendix~\ref{app:B}, the normalized winding number above can be written in the form
\begin{equation}
    w= \sum_{s=\pm } \frac{1}{4\pi \mathrm{i}} \oint \frac{1}{z_s+\mu_0} dz_s , \label{statwinding}
\end{equation}
where $z_\pm = 2 J_0 \cos(k) \pm \mathrm{i} 2\Delta_0 \sin(k)$. Using residue theorem, it then follows that the integration is only non-zero if $\mu_0$ lies inside the loop enclosed by $z_\pm$ in the complex plane, which occurs whenever $|2J_0|>|\mu_0|$. Therefore, at $|2J_0|<|\mu_0|$, $w=0$ and consequently the system is topologically trivial. At $|2J_0|>|\mu_0|$, $w=\pm 1$, rendering the system topologically nontrivial with a pair of gMZMs. This analysis thus confirms the topological origin of the above obtained gMZMs and is summarized in the phase diagram of Fig.~\ref{fig:mzphase}(a). \RB{It is worth mentioning that the definition of $w$ above takes into account all possible values of the conserved quantity $\hat{M}$. Consequently, the calculated $\mathbb{Z}$ invariant and the corresponding phase diagram in Fig.~\ref{fig:mzphase}(a) completely describe the whole system and are not restricted to a specific value of $M$.} 

Finally, to verify that gMZMs are well-gapped from the rest of the bulk energy excitations, we numerically plot the full energy excitation spectrum of our number-conserving system in Fig.~\ref{fig:mzphase}(b). \RB{There, each data point implies the existence of an energy excitation operator $\Gamma_E$ which maps any energy eigenstate $|\epsilon \rangle$ to some $|\epsilon +E\rangle $ that is shifted in energy by $E$. In particular, zero energy excitation solutions correspond to the presence of gMZMs. In the present $E_c=0$ case, we find that such zero solutions, if they exist, are infinitely degenerate instead of being twofold degenerate as in their mean-field counterpart. The reason for this is because in the number conserving setting, a gMZM could map any energy eigenstate $|\epsilon \rangle$ corresponding to a value of $M$ to another eigenstate of the same energy but of $M+(2n+1)$ value for any $n\in\mathbb{Z}$, thus generalizing the ability of an MZM in the mean-field model to connect two degenerate energy eigenstates with opposite parity eigenvalues.} Finally, note that at $\mu=0$, we find that the gap between the gMZMs and other bulk modes is $J$, which agrees with the finding of Ref.~\cite{Lapa2020}. 

\subsection{Time-periodic case}
\label{tpcase}

The mean-field version of Eq.~(\ref{ncons}) in its general time-periodic form has been extensively studied since the last decade, the main interest of which lies in the presence of MPMs that have no static counterparts \cite{FMF1,FMF2,FMF3,FMF6,FMF7}. The advantages of utilizing MPMs for quantum computation have further been uncovered in recent years \cite{FMF5,RG,RG2020b,Matthies2022,FMF4,Bauer2018}. Due to the lack of existing studies, the fate of these MPMs in the number-conserving description of periodically driven topological superconductors has remained unknown and will be the subject of this section. 

For simplicity, we take a binary drive throughout this work under which $\mu(t)=0$, $J(t)=J$, and $\Delta(t)=\Delta$ in the first half of the period, whilst $\mu(t)=\mu$, $J(t)=\Delta(t)=0$ in the second half of the period. We further denote $T$ as the driving period and work in units where $\hbar=1$. In such a time-periodic system, we define the Floquet operator as the one-period time evolution operator
\begin{equation}
    U_T = \mathcal{T} \exp\left(-\mathrm{i} \int_0^T H(t) dt\right) ,
\end{equation}
where $\mathcal{T}$ is the time-ordering operator. Any eigenvalue of $U_T$ can be written in the form $e^{-\mathrm{i} \varepsilon T}$, where $\varepsilon$ is termed the quasienergy due to its analogy with the notion of energy in static systems. The eigenstates of $U_T$ are consequently also termed the quasienergy eigenstates \cite{Shirley1965,Sambe1973}.

We may write the system's Floquet operator in the BdG form as
\begin{equation}
    U_T =\exp\left(-\mathrm{i} \frac{1}{2} \psi^\dagger \mathcal{H}_{\rm eff} T \psi\right) ,
\end{equation}
where the effective BdG Hamiltonian can be found from the BdG Floquet operator
\begin{eqnarray}
    \mathcal{U}_{T} &\equiv & e^{-\mathrm{i} \mathcal{H}_{\rm eff} T} \nonumber \\ 
    &=& e^{-\mathrm{i} \mu \sigma_z \eta_0 T/2} e^{-\mathrm{i} \left[ J \sigma_z \eta_x +\mathrm{i} \Delta  \left(e^{-\mathrm{i} \hat{\phi}} \sigma_+ - e^{\mathrm{i} \hat{\phi}} \sigma_- \right) \eta_y\right] T/2 } . \nonumber \\ \label{nconsu}
\end{eqnarray}
In this case, the quasienergies of the BdG Floquet operator $\mathcal{U}_{T}$ form the quasienergy excitation spectrum of the system. \RB{In particular, a quasienergy excitation solution $\varepsilon$ implies the presence of an operator $\Gamma_\varepsilon$ which maps any quasienergy eigenstate $|\epsilon\rangle$ to some $|\epsilon+\varepsilon\rangle$ that is shifted in quasienergy by $\varepsilon$.} In analogy to its mean-field counterpart, we expect gMZMs to emerge as edge-localized zero quasienergy excitations, whereas the number-conserving analogues of MPMs are expected to instead arise as $\pi/T$ quasienergy excitation solutions.

Under PBC, the system's Floquet operator can further be written as  
\begin{equation}
    U_T =\exp\left(-\mathrm{i} \sum_k \frac{1}{2} \psi_k^\dagger \mathcal{H}_{k,\rm eff} T \psi_k\right) ,
\end{equation}
where the momentum space effective BdG Hamiltonian $\mathcal{H}_{k,\rm eff}$ can be found from the corresponding momentum space BdG Floquet operator, which in the symmetric time frame \cite{Asboth2014,RG2020b} takes the form 
\begin{eqnarray}
    \mathcal{U}_{k,\rm sym} &=& e^{-\mathrm{i} \mu \sigma_z \frac{T}{4}} e^{-\mathrm{i} \left[ 2J \cos(k) \sigma_z - 2\Delta \sin(k)  \left(e^{-\mathrm{i} \hat{\phi}} \sigma_+ - e^{\mathrm{i} \hat{\phi}} \sigma_- \right) \right] \frac{T}{2} }   \nonumber \\
    && \times e^{-\mathrm{i} \mu \sigma_z \frac{T}{4}} \label{usym}
\end{eqnarray}
It is easily verified that the system respects particle-hole, two chiral, and two time-reversal symmetries under the same operators defined in the previous section. In the time-periodic setting, these symmetries read $\mathcal{P} \mathcal{H}_{k,t,\rm sym} \mathcal{P}^\dagger = -\mathcal{H}_{-k,t,\rm sym}$, $\mathcal{C}_1 \mathcal{H}_{k,t,\rm sym} \mathcal{C}_1^\dagger = \mathcal{C}_2 \mathcal{H}_{k,t,\rm sym} \mathcal{C}_2^\dagger = -\mathcal{H}_{k,T-t,\rm sym}$, and $\mathcal{T}_1 \mathcal{H}_{k,t,\rm sym} \mathcal{T}_1^\dagger = \mathcal{T}_2 \mathcal{H}_{k,t,\rm sym} \mathcal{T}_2^\dagger = \mathcal{H}_{-k,T-t,\rm sym}$, where $\mathcal{H}_{k,t,\rm sym}$ is the time-periodic BdG Hamiltonian that generates $\mathcal{U}_{k,\rm sym}$ over one period. The system thus also belongs to the BDI class in the classification of Floquet topological phases \cite{Harper2020}, which is characterized by a $\mathbb{Z}\times \mathbb{Z}$ invariant.   

As detailed in Appendix~{\ref{app:C}}, two types of normalized winding number invariants $w_0$ and $w_\pi$ could be defined \RB{from the half-period BdG time-evolution operator in the symmetric time frame and in the basis where $\mathcal{C}_1$ is diagonal.} At $J=\Delta$, these invariants are found as (see Appendix~\ref{app:C})
\begin{eqnarray}
    w_0&=& \sum_{s=\pm}\frac{1}{4\pi \mathrm{i}} \oint \frac{1}{z_{s,0} +\cos(JT/2)\sin(\mu T/4)} dz_{s,0} \;, \nonumber \\
    && \\
    w_\pi &=& \sum_{s=\pm}\frac{1}{4\pi \mathrm{i}} \oint \frac{1}{z_{s,\pi} +\cos(JT/2)\cos(\mu T/4)} dz_{s,\pi} \, \nonumber \\
\end{eqnarray}
which respectively determine the number of pairs of gMZMs and generalized MPMs (gMPMs). There, $z_{\pm,0} = e^{\pm \mathrm{i} k} \cos(\mu T/4)\sin(JT/2)$ and $z_{\pm,\pi} = e^{\pm \mathrm{i} k} \sin(\mu T/4)\sin(JT/2)$. Using residue theorem, it then follows that $w_0=\pm 1$ if $|\tan(\mu T/4)|<|\tan(JT/2)|$ and $w_0=0$ otherwise, whilst $w_\pi=\pm 1$ if $|\tan(\mu T/4)|>|\cot(JT/2)|$ and $w_\pi=0$ otherwise. In the regime $(\mu T , 2JT) = (0,2\pi] \times (0,2\pi]$, which will be assumed throughout the rest of this paper, gMZMs (gMPMs) exist whenever $\mu < 2J$ ($\mu>2\pi/T -2J$). In particular, gMZMs and  gMPMs can both coexist if both conditions are satisfied simultaneously. This leads to a phase diagram shown in Fig.~\ref{fig:mzphase}(c). It is worth noting that such a phase diagram is identical to that of the mean-field $p$-wave superconductor \cite{FMF7}, further highlighting the correspondence between the gMZMs and gMPMs in our number-conserving system and their true Majorana mode counterparts. \RB{Moreover, as the definitions of $w_0$ and $w_\pi$ in Appendix~\ref{app:D} involve all values of $M$, the above conclusion holds generally and not only at specific values of $M$.} In the following, we will further present the exact of expressions for gMZMs and/or gMPMs at three special points.

{\bf (i) $\mu=0$ case:} Equation~(\ref{nconsu}) then becomes equivalent to the time-evolution under Eq.~(\ref{nconsstat}), which is known to support gMZMs of the form $\hat{\Gamma}_{A,1}$ and $\hat{\Gamma}_{B,L}$ as defined in Sec.~\ref{ticase}. 

{\bf (ii) $\mu T = 2 \pi$ case:} We may first write
\begin{equation}
    U_T = e^{-\sum_j \mathrm{i} \pi c_j^\dagger c_j } U_s,
\end{equation}
where $U_s = e^{-\mathrm{i} H_{\rm stat} T/2}$ is the time evolution under Eq.~(\ref{nconsstat}). It then follows that $[U_s,\hat{\Gamma}_{A,1}] = [U_s,\hat{\Gamma}_{B,L}] =0$ as $\hat{\Gamma}_{A,1}$ and $\hat{\Gamma}_{B,L}$ are gMZMs of $H_{\rm stat}$. Meanwhile, by using the identity 
\begin{eqnarray}
    e^{\sum_j \mathrm{i} \theta c_j^\dagger c_j } c_j \RB{e^{-\sum_j \mathrm{i} \theta c_j^\dagger c_j }} &=& e^{-\mathrm{i} \theta} c_j \;,  \label{idenon} 
\end{eqnarray}
it is easily checked that $e^{\sum_j \mathrm{i} \pi c_j^\dagger c_j } \hat{\Gamma}_{A,1} e^{-\sum_j \mathrm{i} \pi c_j^\dagger c_j } = -\hat{\Gamma}_{A,1} $ and $e^{\sum_j \mathrm{i} \pi c_j^\dagger c_j } \hat{\Gamma}_{B,L} e^{-\sum_j \mathrm{i} \pi c_j^\dagger c_j } = -\hat{\Gamma}_{B,L} $. Consequently, $\hat{\Gamma}_{A,1}$ and $\hat{\Gamma}_{B,L}$ are both  gMPMs of the system with $U_T^\dagger \hat{\Gamma}_{A,1} U_T = -\hat{\Gamma}_{A,1}$ and $U_T^\dagger \hat{\Gamma}_{B,L} U_T = -\hat{\Gamma}_{B,L}$.

{\bf (iii) $\mu T = JT = \pi$ case:} The Floquet operator reads
\begin{equation}
    U_T = e^{-\sum_j \mathrm{i} \pi/2 c_j^\dagger c_j } e^{-\sum_j  \pi/2 \hat{\Gamma}_{A,j+1}^\dagger \hat{\Gamma}_{B,j} } , 
\end{equation}
where the operators $\hat{\Gamma}_{A,j}$ and $\hat{\Gamma}_{B,j}$ are as defined in Sec.~\ref{ticase}. As shown in Appendix~\ref{app:D}, the following identities hold,
\RB{\begin{eqnarray}
    u_\theta \hat{\Gamma}_{A,\ell+1} u_\theta^\dagger &=& \cos(2\theta) \hat{\Gamma}_{A,\ell+1} - \sin(2\theta) \hat{\Gamma}_{B,\ell} \;, \nonumber \\
    u_\theta \hat{\Gamma}_{B,\ell}^\dagger u_\theta^\dagger &=& \cos(2\theta) \hat{\Gamma}_{B,\ell}^\dagger  +  \sin(2\theta) \hat{\Gamma}_{A,\ell+1}^\dagger , \label{idgam}
\end{eqnarray}}
where \RB{$u_\theta=e^{\sum_j  \theta \hat{\Gamma}_{A,j+1}^\dagger \hat{\Gamma}_{B,j} }$}, $\ell=1,\cdots, L-1$ ($\hat{\Gamma}_{A,1}$ and $\hat{\Gamma}_{B,L}$ commute with \RB{$u_\theta$}). Together with Eq.~(\ref{idenon}), this leads to
\begin{eqnarray}
    U_T^\dagger \hat{\Gamma}_{A,1} U_T &=& - \hat{\Gamma}_{B,1} , \nonumber \\
    U_T^\dagger \hat{\Gamma}_{B,1} U_T &=& - \hat{\Gamma}_{A,1} .
\end{eqnarray}
In particular, $\hat{\Gamma}_{\pm,1} \equiv \hat{\Gamma}_{A,1} \pm \hat{\Gamma}_{B,1}$ satisfy $U_T^\dagger \hat{\Gamma}_{\pm,1} U_T = \mp \hat{\Gamma}_{\pm,1}$. That is, $\hat{\Gamma}_{+,1}$ and $\hat{\Gamma}_{-,1}$ are respectively the system's  gMPMs and gMZMs localized near the left end. In a similar fashion, it can be shown that $\hat{\Gamma}_{+,L}$ and $\hat{\Gamma}_{-,L}$ are respectively the system's gMZMs and  gMPMs localized near the right end.

Finally, to support our analytical results above, we plot in Fig.~\ref{fig:mzphase}(d) the quasienergy excitation spectra of our system at two fixed parameter $J=\Delta$ values, one of which supports a regime with coexisting gMZMs and  gMPMs. As expected, the presence of gMZMs and  gMPMs agrees with the phase diagram of Fig.~\ref{fig:mzphase}(c). Moreover, such gMZMs and  gMPMs are well-gapped from the bulk modes.

\subsection{Effect of finite charging energy}
\label{finEc}

It is convenient to write the charging energy term as $E_c(2\hat{n} -2n_c)^2= \psi^\dagger \RB{\frac{E_c}{L}}(2\hat{n} -2n_c)^2 \psi$, where $\psi$ is the Nambu vector defined in Sec.~\ref{ticase}. Indeed, by using the fermionic anticommutation relation $\left\lbrace c_j^\dagger , c_j\right\rbrace =1$, it is easily verified that \RB{$\psi^\dagger \psi = L$}, thus justifying the above equality. In this case, Eq.~(\ref{ncons}) can be written in the BdG-like form as
\begin{eqnarray}
    H(t) &=& \frac{1}{2} \psi^\dagger \mathcal{H}(t) \psi \;, \nonumber \\
    \mathcal{H}(t) &=& J(t) \sigma_z \eta_x +i \Delta (t)  (e^{-\mathrm{i} \hat{\phi}} \sigma_+ - e^{\mathrm{i} \hat{\phi}} \sigma_- ) \eta_y + \mu(t) \sigma_z \eta_0 \nonumber \\
    && + \RB{\frac{2E_c}{L}}(2\hat{n} -2n_c)^2 \sigma_0 \eta_0 ,
\end{eqnarray}
where $\sigma_0$ is the $2\times 2$ identity matrix and the other matrices are the same as those defined in Sec.~\ref{ticase}.

\begin{center}
\begin{figure*}
    \includegraphics[scale=0.45]{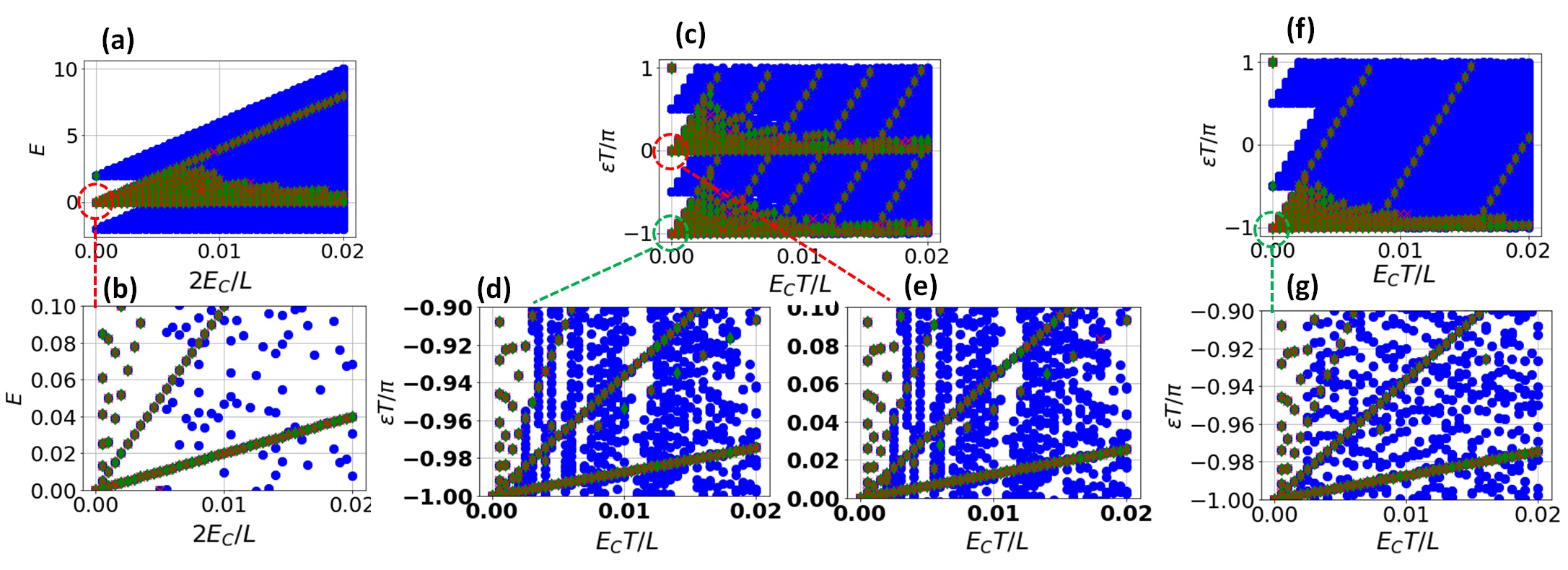}
    \caption{(a) The energy excitation spectrum of the static number-conserving $p$-wave superconductor in the presence of charging energy at $J=\Delta=1$, $\mu=0$. Panel (b) is the zoomed-in version of panel (a) in the vicinity of zero energy. (c,f) The quasienergy excitation spectrum of the periodically driven $p$-wave superconductor in the presence of charging energy, under parameter values which support (c) both gMZMs and  gMPMs at $JT=\Delta T= \mu T =\pi$, and $n_c=0$, and (d)  gMPMs only at $JT=\Delta T= \pi/2$, $\mu T =2\pi$. Panels (d) and (g) are respectively the zoomed-in version of panels (c) and (f) in the vicinity of $-\pi/T$ quasienergy. Panel (e) is the zoomed-in version of panel (c) in the vicinity of zero quasienergy. In all panels, green and red marks respectively represent solutions that are localized near the left and right ends of the system, the system size is taken as $L=15$, $n_c=0$, and the Cooper-pairs degree of freedom is truncated at $n_{\rm min} = -10$ and $n_{\rm max} = 10$.}
    \label{fig:mzpmec}
\end{figure*}    
\end{center}

By first focusing on the time-independent case with $J(t)=J_0$, $\Delta(t)=\Delta_0$, and $\mu(t)=\mu_0$, it follows that the presence of the charging energy term breaks the two chiral symmetries $\mathcal{C}_1$ and $\mathcal{C}_2$ identified in Sec.~\ref{ticase}. Consequently, it is expected that the previously identified gMZMs, if they remain present, are generally no longer pinnned at zero energy. Indeed, at $\mu_0=0$ and $\Delta_0=J_0$, the energy shift associated with the gMZM solutions $\hat{\Gamma}_{A,1}$ and $\hat{\Gamma}_{B,L}$ can be estimated via applying the first-order perturbation theory in $E_c$ with respect to $|A,1\rangle \equiv (e^{\mathrm{i} \hat{\phi}},0,\cdots,1,\cdots,0)/\sqrt{2}$ and $|B, L\rangle \equiv \mathrm{i} (0,\cdots,e^{\mathrm{i} \hat{\phi}},0,\cdots ,-1)/\sqrt{2}$ states in the Nambu basis. In both cases, we find
\begin{eqnarray}
    \Delta E_n &\approx & \langle E_c(2\hat{n} -2n_c)^2 \sigma_0 \eta_0 \rangle \nonumber \\
    &=& \frac{4 E_c}{L} \left[ (n-n_c)^2 + (n -(n_c-1))^2\right] , \label{firstodstat}
\end{eqnarray}
where $n$ is the number of Cooper pairs. In particular, the dependence of the energy shift on $n$ implies that the charging energy term breaks the infinite degeneracy of gMZMs \RB{discussed in Sec.~\ref{ticase}} in the Cooper-pairs subspace at zero energy into two-fold degeneracy at $\Delta E_n$, each comprises of a left-edge localized and a right-edge localized generalized Majorana modes. To support these results, we numerically compute the energy spectrum of the BdG-like Hamiltonian in Fig.~\ref{fig:mzpmec}(a,b) as a function of $E_c$ without making a first-order approximation. In particular, the linear relation between the energy excitations of gMZMs and $E_c$ shown in Fig.~\ref{fig:mzpmec}(b) agrees very well with Eq.~(\ref{firstodstat}).  

Moving on to the time-periodic case, we take the same driving scheme as that in Sec.~\ref{tpcase}. The system's BdG Floquet operator can be written explicitly as
\begin{eqnarray}
    \mathcal{U}_{T} &=& e^{-\mathrm{i} \left(\mu \sigma_z + \frac{2E_c}{L}(2\hat{n} -2n_c)^2 \sigma_0 \right) \eta_0 T/2} \nonumber \\
    &\times & e^{-\mathrm{i} \left[ J \sigma_z \eta_x +\mathrm{i} \Delta  \left(e^{-\mathrm{i} \hat{\phi}} \sigma_+ - e^{\mathrm{i} \hat{\phi}} \sigma_- \right) \eta_y + \frac{2E_c}{L}(2\hat{n} -2n_c)^2 \sigma_0 \eta_0 \right] T/2 } . \nonumber \\
\end{eqnarray}
Similarly to the time-independent case above, the presence of the charging energy term breaks the two chiral symmetries $\mathcal{C}_1$ and $\mathcal{C}_2$ of Sec.~\ref{tpcase}, thus shifting the existing gMZMs and  gMPMs in quasienergy away from zero and $\pm \pi/T$ respectively. 

At $JT=\Delta T=\mu T=\pi$, our analysis in Sec.~\ref{tpcase} shows that the system supports a pair of  gMPMs as $\hat{\Gamma}_{+,1}$ and $\hat{\Gamma}_{-,L}$ and a pair of gMZMs as $\hat{\Gamma}_{-,1}$ and $\hat{\Gamma}_{+,L}$. The shift in quasienergy of the above gMZMs and  gMPMs away from zero and $\pm \pi/T$ respectively can be obtained perturbatively up to first order in $E_c$ with respect to the BdG states $|\pm ,1\rangle = \frac{e^{\pm \mathrm{i} \pi/4}}{\sqrt{2}} (1,0,\cdots,\mp \mathrm{i} e^{-\mathrm{i} \hat{\phi}},\cdots,0)$ and $|\pm ,L\rangle = \frac{e^{\pm \mathrm{i} \pi/4}}{\sqrt{2}} (0,0,\cdots,1, 0, \cdots ,\mp \mathrm{i} e^{-\mathrm{i} \hat{\phi}})$. In all cases, we find that 
\begin{equation}
    \Delta \varepsilon_n^{(gMZM)} = \Delta \varepsilon_n^{(gMPM)} = \Delta E_n \;\;{\rm mod}\;\; 2\pi/T ,  
\end{equation}
where $\varepsilon_n^{(gMZM)}$ ($\varepsilon_n^{(gMPM)}$) is the shift in quasienergy of the gMZMs (gMPMs), $\Delta E_n$ is given by Eq.~(\ref{firstodstat}), and the modulus of $2\pi/T$ exists to account for the fact that quasienergy is only defined within $\left( -\pi/T , \pi/T\right]$. 

At $J=\Delta$ and $\mu T=2\pi$, our analysis in Sec.~\ref{tpcase} shows that the system only supports a pair of  gMPMs as $\hat{\Gamma}_{A,1}$ and $\hat{\Gamma}_{B,L}$, which respectively correspond to the BdG states $|A,1\rangle$ and $|B,L\rangle$. Up to first order in $E_c$, both  gMPMs then acquire a shift in quasienergy away from $\pm \pi/T$ by $\Delta \varepsilon_n = \Delta E_n \;\;{\rm mod}\;\; 2\pi/T$, where $\Delta E_n$ is given by Eq.~(\ref{firstodstat}). This result is confirmed numerically in Fig.~\ref{fig:mzpmec}(f,g).      

\section{Quantum computing prospects of gMZMs and gMPMs}
\label{disc}

\subsection{Robustness against spatial disorders}
\label{discdis}

Any realistic system is inherently imperfect. Spatial disorders are some of the most ubiquitous sources of imperfection that could result, e.g., from our inability to fine tune system parameters and/or from the presence of impurities in the systems. It is widely-known that MZMs and MPMs in the mean-field description of periodically driven topological superconductors are robust against such disorders. This is one of the most important features that makes Majorana-based qubits incredibly attractive. Whether gMZMs and gMPMs in the number-conserving periodically driven topological superconductors also enjoy the same robustness against spatial disorders has however remained unexplored, thus obscuring their potential quantum computing advantages. \RB{It is natural to expect that gMZMs and gMPMs should be similarly resilient against disorders due to the presence of protecting topological invariants, provided such disorders do not accidentally close the bulk gap. If such an accidental bulk gap closing occurs, a topological phase transition may take place which ultimately leads to the destruction of gMZMs and gMPMs. To rule out this possibility, studying the effect of disorder on the above obtained gMZMs and gMPMs is of great importance.}

\begin{center}
\begin{figure}
    \includegraphics[scale=0.43]{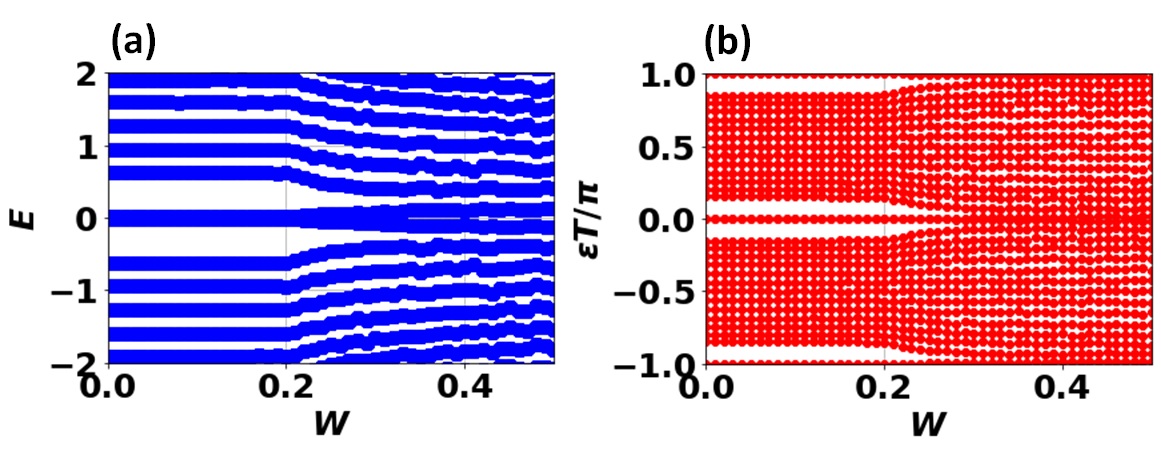}
    \caption{(a) The disorder averaged energy excitation spectrum of the static system in Sec.~\ref{ticase} at $\overline{J}_0=\overline{\Delta}_0=0.4 \pi$ and $\overline{\mu}_0= 0.5\pi$. (b) The disorder averaged quasienergy excitation spectrum of the periodically driven system in Sec.~\ref{tpcase} at $\overline{J} T=\overline{\Delta} T=0.8 \pi$ and $\overline{\mu} T= \pi$. All data points are taken at $E_c=0$, $L=15$, by truncating the Cooper-pairs degree of freedom at $n_{\rm min}=-10$ and $n_{\rm max}=10$, and by averaging over $100$ disorder realizations.}
    \label{fig:disorder}
\end{figure}    
\end{center}

In view of the above, we explicitly investigate the presence of spatial disorders in our system of Eq.~(\ref{ncons}) by replacing the the system parameters $J(t)\rightarrow J_j(t)$, $\Delta(t)\rightarrow \Delta_j(t)$, and $\mu(t)\rightarrow \mu_j(t)$ (we set $E_c=0$ for simplicity). In Fig.~\ref{fig:disorder}(a), we plot the disorder averaged energy excitation spectrum of our static number-conserving $p$-wave superconductor described in Sec.~\ref{ticase}. There, the parameters $J_{0,j}$, $\Delta_{0,j}$, and $\mu_{0,j}$ are drawn uniformly from $[\overline{P}-W\overline{P},\overline{P}+W\overline{P}]$, where $P\in \left\lbrace J_{0}, \Delta_0, \mu_0 \right\rbrace$ is the associated system parameter introduced in Sec.~\ref{ticase} and $W$ is the disorder strength. In Fig.~\ref{fig:disorder}(b), We further plot the disorder averaged quasienergy spectrum of our periodically driven system described in Sec.~\ref{tpcase}. There the parameters are drawn uniformly from $[\overline{P}-W\overline{P},\overline{P}+W\overline{P}]$, where $P\in \left\lbrace J, \Delta, \mu \right\rbrace$ is the associated system parameter introduced in Sec.~\ref{tpcase} and $W$ is the disorder strength.  

Our results demonstrate that gMZMs and gMPMs are indeed robust against small to moderate spatial disorders. In particular, not only do they remain present at finite disorders, but they are also pinned at their expected quasienergies ($0$ for gMZMs and $\pi/T$ for gMPMs). These modes only disappear or lose their topological protection at a very large disorder, which results in a bulk gap closing. However, as shown in Fig.~\ref{fig:disorder}, this only occurs at disorder strengths $\gtrapprox 25\%$ of the system parameter values, which are easily avoided in experiments.

\subsection{Robustness against finite charging energy} 
\label{discchar}
%The most attractive aspect of Majorana modes is their ability to encode qubits nonlocally via their corresponding parity operator. 

In Sec.~\ref{finEc}, we have shown that at finite charging energy, the generalized Majorana modes of number-conserving $p$-wave superconductors remain present, but they are shifted from zero energy (zero or $\pi/T$ quasienergy) in the static (time-periodic) case. A natural question then arises in regard to the impact of such a shift in energy/quasienergy to their ability to store and process qubits. 
 %Investigating how the generalized Majorana parities are potentially affected by such a shift in energy/quasienergy of the generalized Majorana modes thus forms an extremely important subject to assess the suitability of number-conserving topological superconductors for quantum computing applications.

To address the above question, we analyze the generalized Majorana parity $P_{1,2} = \mathrm{i} \hat{\Gamma}_1^\dagger \hat{\Gamma}_2$ associated with two generalized Majorana modes $\hat{\Gamma}_1$ and $\hat{\Gamma}_2$. In the static case when $\hat{\Gamma}_1$ and $\hat{\Gamma}_2$ are necessarily gMZMs, such a parity operator satisfies $[H,P_{1,2}]=0$, which implies that all eigenenergies of $H$ can be arranged according to the $\pm 1$ eigenvalues of $P_{1,2}$. In the absence of the charging energy, we additionally have that $[H,\hat{\Gamma}_1]=[H,\hat{\Gamma}_2]=0$. Due to the anti-commutation relation
\begin{equation}
  \left\lbrace P_{1,2} , \hat{\Gamma}_1 \right\rbrace = \left\lbrace P_{1,2} , \hat{\Gamma}_2 \right\rbrace = 0 , \label{antcom}  
\end{equation}
   the operators $P_{1,2}$, $\hat{\Gamma}_1$, and $\hat{\Gamma}_2$ cannot all be simultaneously diagonalized. Therefore, given an energy eigenstate $|E,+\rangle$ that satisfies $H |E,+\rangle = E |E,+\rangle$ and $P_{1,2} |E,+\rangle = + |E,+\rangle$, the corresponding state $\hat{\Gamma}_1 |E,+\rangle$ is easily shown to satisfy $H \hat{\Gamma}_1 |E,+\rangle = E \hat{\Gamma}_1 |E,+\rangle$ and $P_{1,2} 
 \hat{\Gamma}_1|E,+\rangle = -\hat{\Gamma}_1 |E,+\rangle$. That is, $\hat{\Gamma}_1 |E,+\rangle$ represents another energy eigenstate of the same $E$ but of opposite $-1$ parity eigenvalue. As this is true for any energy eigenstate of $H$, it follows that all energy eigenvalues of $H$ are at least two-fold degenerate. Similar to its mean-field counterpart, any of these degenerate energy subspaces may in turn be utilized to store a qubit. In particular, as the degenerate subspace is characterized by an inherently nonlocal operator $P_{1,2}$, the encoded qubit is expected to be very robust.

   At finite charging energy, $H$ no longer commutes with $\hat{\Gamma}_1$ and $\hat{\Gamma}_2$. However, if $H$ still manages to commute with $P_{1,2}$, its energy eigenvalues could still be labelled by the eigenvalue of $P_{1,2}$. Moreover, as the anticommutation relation Eq.~(\ref{antcom}) still holds, it follows that $\hat{\Gamma}_1 |E,+\rangle = |E',-\rangle $, but with $E\neq E'$ in general. Therefore, one may in principle encode a qubit in the (nondegenerate) subspace spanned by $|E,+\rangle$ and $ |E',-\rangle$, which is also expected to enjoy the topological robustness due to the nonlocality of the parity operator $P_{1,2}$. 

   While a nonlocal qubit could still be encoded with two gMZMs at finite charging energy, the use of nondegenerate subspace might make the resulting qubit more prone to dephasing. Fortunately, one may recover the degeneracy of the nonlocal qubit subspace by utilizing at least four gMZMs, e.g., $\hat{\Gamma}_1$, $\hat{\Gamma}_2$, $\hat{\Gamma}_3$, and $\hat{\Gamma}_4$, and requiring that $[H,P_{1,2}]=[H,P_{1,3}]=0$. Indeed, as $\left\lbrace P_{1,2} , P_{1,3} \right\rbrace = 0$, the three operators $H$, $P_{1,2}$, and $P_{1,3}$ cannot be simultaneously diagonalized. By choosing to simultaneously diagonalize $H$ and $P_{1,2}$, the remaining $P_{1,3}$ operator then guarantees that all eigenenergies of $H$ are at least two-fold degenerate, as $P_{1,3} |E,\pm \rangle \propto |E,\mp\rangle$, where $|E,\pm \rangle$ is the energy $E$ eigenstate of $H$ associated with $P_{1,2}=\pm 1$ eigenvalue. In this case, $P_{1,2}$ and $P_{1,3}$ serve as the effective Pauli matrices $\sigma_z$ and $\sigma_x$ in the degenerate qubit subspace.

The above argument demonstrates that the preservation of zero commutator between the system's Hamiltonian and the parity operators is crucial for the ability of gMZMs to form a nonlocal qubit subspace. We will now show that this is indeed the case for our number-conserving $p$-wave superconductors, even at finite charging energy. To this end, we first define the zero energy excitation spectral function $s_0^{(P_{i,j})}$ associated with the parity operator $P_{i,j}$ as \cite{Sreejith2016,BomantaraSen2021,Bomantara2021a,Bomantara2022a}
\begin{eqnarray}
    s_0^{(P_{i,j})} &=& \frac{1}{\mathcal{N}^{(P_{i,j})}} \sum_{|E\rangle \in \chi} \int_{-\delta}^\delta S^{(P_{i,j})}\left(|E\rangle, \eta\right) d\eta , \nonumber \\
    \mathcal{N}^{(P_{i,j})} &=& \sum_{|E\rangle \in \chi} \int_{-\infty}^\infty S^{(P_{i,j})}\left(|E\rangle, \eta\right) d\eta\nonumber \\
    S^{(P_{i,j})}\left(|E\rangle, \eta\right) &=& \sum_{|E'\rangle \in \mathcal{H}} \delta(E-E'-\eta) |\langle E' | P_{i,j} | E\rangle |^2 ,\nonumber \\ 
    \label{statspec}
\end{eqnarray}
where $\delta$ is a sufficiently small number in units of energy, $\mathcal{H}$ is the set of all eigenstates of the system's Hamiltonian, and $\chi \in \mathcal{H}$ contains a smaller number (taken as $32$ in this manuscript) of the system's randomly chosen energy eigenstates. 

\begin{center}
\begin{figure}
    \includegraphics[scale=0.41]{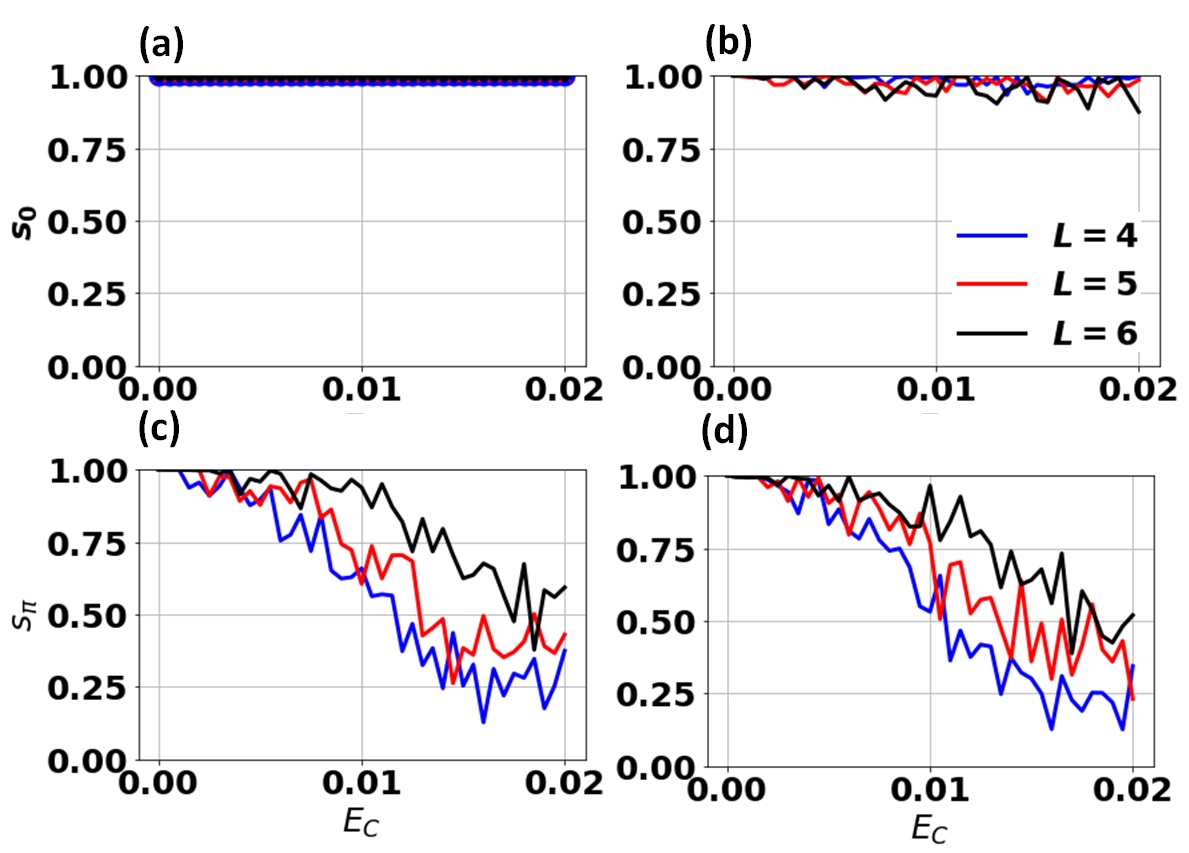}
    \caption{(a) The spectral function $s_0^{\left(P_{L,R}\right)}$ for the static number-conserving $p$-wave superconductor of Sec.~\ref{ticase} at $J_0=\Delta_0 =1$ and $\mu_0=0$. (b,c,d) The spectral functions for the periodically driven number-conserving $p$-wave superconductor of Sec.~\ref{tpcase} corresponding to (b) $s_0^{\left(P_{(\pi,L),(\pi,R)}\right)}$, (c) $s_\pi^{\left(P_{(\pi,L),(0,L)}\right)}$, (d) $s_\pi^{\left(P_{(0,L),(\pi,R)}\right)}$, at $J T=\Delta T =\mu T=\pi$. In all panels, $n_c=0$ and the Cooper-pairs degree of freedom is truncated at $n_{\rm min} = -8$ and $n_{\rm max} = 8$.}
    \label{fig:charg}
\end{figure}    
\end{center}

Note that $s_0^{(P_{i,j})}=1$ if $[P_{i,j},H]=0$. In general, $P_{i,j}$ may not commute exactly with $H$ due to the finite system size. However, it is expected that $s_0^{(P_{i,j})}$ is still close to unity as long as $P_{i,j}$ approximately commutes with $H$. For our static number-conserving $p$-wave superconductor of Sec.~\ref{ticase}, we define the relevant parity operator as $P_{L,R} = \mathrm{i} \hat{\Gamma}_L^\dagger \hat{\Gamma}_R$. Here, the gMZMs $\hat{\Gamma}_L$ and $\hat{\Gamma}_R$ at finite charging energy are obtained by adiabatically evolving Eq.~(\ref{exactMM}) as $E_c$ is slowly increased from $0$. In Fig.~\ref{fig:charg}, it is indeed verified that $s_0^{\left(P_{L,R}\right)}$ is very close to $1$ for all values of $E_c$ under consideration. This result thus demonstrates the robustness of the nonlocal qubit subspace in the presence of the charging energy even though its gMZM constituents are shifted in energy. 

A similar analysis will now be repeated to demonstrate the robustness of the nonlocal qubit subspace formed by the gMZMs and gMPMs of our periodically driven $p$-wave superconductor of Sec.~\ref{tpcase} at a finite charging energy. To this end, we first note that Eq.~(\ref{antcom}) is satisfied regardless of whether $\hat{\Gamma}_1$ and $\hat{\Gamma}_2$ are gMZMs or gMPMs. By further requiring that $[U_T,P_{1,2}]=0$, the same argument above then implies that a qubit subspace spanned by some (generically nondegenerate) quasienergy states $|\varepsilon,+\rangle$ and $|\varepsilon',-\rangle \propto \hat{\Gamma}_1 |\varepsilon,+\rangle $. 

A scenario of interest in our periodically driven $p$-wave superconductor involves the presence of two gMZMs $\hat{\Gamma}_{0,L}$ and $\hat{\Gamma}_{0,R}$ and two gMPMs $\hat{\Gamma}_{\pi,L}$ and $\hat{\Gamma}_{\pi,R}$ (see the gray region of Fig.~\ref{fig:mzphase}(c)). In this case, two types of generalized Majorana parity operators can be defined, e.g., $P_{(\pi,L),(\pi,R)}=\mathrm{i} \hat{\Gamma}_{\pi,L}^\dagger \hat{\Gamma}_{\pi,R}$ and $P_{(0,L),(\pi,L)}=\mathrm{i} \hat{\Gamma}_{0,L}^\dagger \hat{\Gamma}_{\pi,L}$, which satisfy $[U_T,P_{(\pi,L),(\pi,R)}]=0$ and $\left\lbrace U_T,P_{(0,L),(\pi,L)}\right\rbrace=0$. The additional parity operator which anticommutes with the Floquet operator implies that the quasienergies $\varepsilon$ and $\varepsilon'$ spanning the qubit subspace have a difference of $\pi/T$. Consequently, the qubit subspace is effectively degenerate when viewed by $U_T^2$. It is then expected that performing a stroboscopic quantum computing--in which gate operations and measurements are executed at discrete times that are integer multiples of $2T$--on such a qubit subspace will yield the same advantage as that on a typical static degenerate qubit subspace. Here, the use of periodic driving has the advantage of reducing the physical resource needed to support four generalized Majorana modes, i.e., due to the presence of gMPMs.    

In view of the above, we will now numerically check that the appropriate generalized parity operators tend to commute/anticommute with the Floquet operator even at finite charging energy. To this end, we now define two sets of spectral functions
\RB{\begin{eqnarray}
    s_0^{(P_{i,j})} &=& \sum_{|\varepsilon \rangle \in \chi} \sum_{n=0,\pm 1} \int_{2\pi n/T -\delta}^{2\pi n/T+\delta}  \frac{S^{(P_{i,j})}(|\varepsilon\rangle,\eta)}{\mathcal{N}^{(P_{i,j})}} d\eta ,  \nonumber \\
    s_\pi^{(P_{i,j})} &=& \sum_{|\varepsilon \rangle \in \chi} \sum_{n=\pm 1} \int_{\pi n /T -\delta}^{\pi n /T+\delta}  \frac{S^{(P_{i,j})}(|\varepsilon\rangle,\eta)}{\mathcal{N}^{(P_{i,j})}} d\eta ,  \label{tpspec}  
\end{eqnarray}}

\noindent where $\delta$ is a sufficiently small number in units of quasienergy, $S^{(P_{i,j})}\left(|\varepsilon \rangle, \eta\right)$ is as defined in Eq.~(\ref{statspec}), but with $\mathcal{H}$ being the set of all quasienergy eigenstates, $\chi \in \mathcal{H}$ contains a smaller number (taken as $32$ in this manuscript) of the system's randomly chosen quasienergy eigenstates, and $\mathcal{N}^{(P_{i,j})} = \sum_{|\varepsilon \rangle \in \chi} \int_{-\pi/T}^{\pi/T} S^{(P_{i,j})}\left(|\varepsilon \rangle, \eta\right) d\eta$ is the normalization constant. 

The difference in the integration limit between Eqs.~(\ref{statspec}) and (\ref{tpspec}) is to account for the fact that the quasienergies are only defined modulo $2\pi/T$. Note that $s_0^{(P_{i,j})}=1$ if $[U_T,P_{i,j}]=0$ and $s_\pi^{(P_{i,j})}=1$ if $\left\lbrace U_T,P_{i,j} \right\rbrace=0$. In Fig.~\ref{fig:charg}(b-d), we plot the appropriate spectral functions for three generalized Majorana parity operators as a function of the charging energy. Observe that the spectral function $s_0$ associated with a parity that is made up of two gMPMs (panel b) remains very close to $1$ for all charging energy values under consideration. A very qualitatively similar profile is obtained for the parity made up of two gMZMs (not shown in the figure). These results show that generalized Majorana parity operators made up of two gMZMs or two gMPMs effectively commute with the Floquet operator as expected, even at finite charging energy. For the two generalized Majorana parity operators made up of one gMZM and one gMPM (panels c and d), the spectral function $s_\pi$ is still finite, albeit not very close to unity. However, it is clearly observed that $s_\pi$ moves closer towards unity with increase in the system size. In the thermodynamic limit, it is then expected that $s_\pi \rightarrow 1$ and, consequently, the associated parities anticommute with $U_T$. These results demonstrate the robustness of the qubit subspace in static and periodically driven number-conserving $p$-wave superconductors in the presence of charging energy.

\subsection{Braiding simulation between a gMZM and a gMPM}
\label{discbraid}

To fully demonstrate the quantum computing abilities of gMZMs and gMPMs in a number-conserving periodically driven topological superconductor, we explicitly design and execute a braiding between a gMZM and a gMPM in our system. To this end, we adapt the braiding scheme developed in \cite{FMF4}, which will be elaborated in the following.

We first consider the special parameter values of (iii) considered in Sec.~\ref{tpcase}, under which our number-conserving periodically driven $p$-wave superconductor supports $\hat{\Gamma}_{+,1}$ and $\hat{\Gamma}_{-,1}$ as a gMZM and a gMPM respectively. In this case, the system's Hamiltonian over one period can be written as
\begin{eqnarray}
    H(t) &=& \begin{cases}
        H_1 \equiv -\mathrm{i} \sum_{j=1}^{L-1} \frac{\pi}{T} \hat{\Gamma}_{A,j+1}^\dagger \hat{\Gamma}_{B,j} & \text{ for } 0<\frac{t}{T}\leq \frac{1}{2} \\
        H_2 \equiv \sum_{j=1}^{L-1} \frac{\pi}{T} c_j^\dagger c_j & \text{ for } \frac{1}{2}<\frac{t}{T}\leq 1
    \end{cases} . \nonumber \\ 
    \label{Hini}
\end{eqnarray}
Our braiding scheme consists of six steps and involves slowly changing $H_1$ at every other period while keeping $H_2$ the same. Specifically, At step $x=1,\cdots,6$ in our scheme, we take $H_{x}(s)= s H_{1,x+1} +(1-s) H_{1,x} $, where $s=\frac{j}{\tau}$ when $2 j T< t\leq 2(j+1)T$, $j=0,1,\cdots , \tau$ so that $t_f= 2(\tau +1) T$ is the total time to complete one step,  
\begin{eqnarray}
    H_{1,1}=H_{1,4}&=&H_{1,7}= H_1 \;, \nonumber \\
    H_{1,2}=H_{1,5} &,& H_{1,3}=H_{1,6} \;,
\end{eqnarray}
and
\begin{eqnarray}
    H_{1,2} &=& -\mathrm{i} \frac{\pi}{T} \left(\hat{\Gamma}_{A,1}^\dagger \hat{\Gamma}_{B,2} - \hat{\Gamma}_{A,3}^\dagger \hat{\Gamma}_{B,2} \right) + H_{1,1} ,\nonumber  \\
    H_{1,3} &=& -\mathrm{i} \frac{\pi}{T} \left(\hat{\Gamma}_{B,1}^\dagger \hat{\Gamma}_{A,2} - \hat{\Gamma}_{A,3}^\dagger \hat{\Gamma}_{A,2} \right) + H_{1,2} .
\end{eqnarray}
In particular, 
\begin{eqnarray}
    H_{1}(s) &=& -\mathrm{i} \frac{\pi}{T} \left[(1-s) \Gamma_{A,3}^\dagger + s \Gamma_{A,1}^\dagger \right] \Gamma_{B,2} + \cdots ,\nonumber \\
    H_2 (s) &=& \mathrm{i} \frac{\pi}{T} \left[(1-s) \Gamma_{B,1}^\dagger + s \Gamma_{A,3}^\dagger \right] \Gamma_{A,2} + \cdots , \nonumber \\
    H_3 (s) &=& -\mathrm{i} \frac{\pi}{T} \left\lbrace (1-s) \left(-\hat{\Gamma}_{A,3}^\dagger \hat{\Gamma}_{A,2} + \hat{\Gamma}_{A,1}^\dagger \hat{\Gamma}_{B,2} \right)\right. \nonumber \\
    && + s \left. \left(\hat{\Gamma}_{A,3}^\dagger \hat{\Gamma}_{B,2} - \hat{\Gamma}_{B,1}^\dagger \hat{\Gamma}_{A,2} \right) \right\rbrace +\cdots  
\end{eqnarray}
$H_4(s)=H_1(s)$, $H_5(s)=H_2(s)$, and $H_6(s)=H_3(s)$. The rationale for varying the adiabatic parameter every other period is to ensure that gMZMs and gMPMs become effectively degenerate from the perspective of $U_T^2$, so that a non-Abelian Berry phase could be accumulated to turn a gMZM into a gMPM and vice versa \cite{FMF4}. 

\begin{center}
\begin{figure*}
    \includegraphics[scale=0.5]{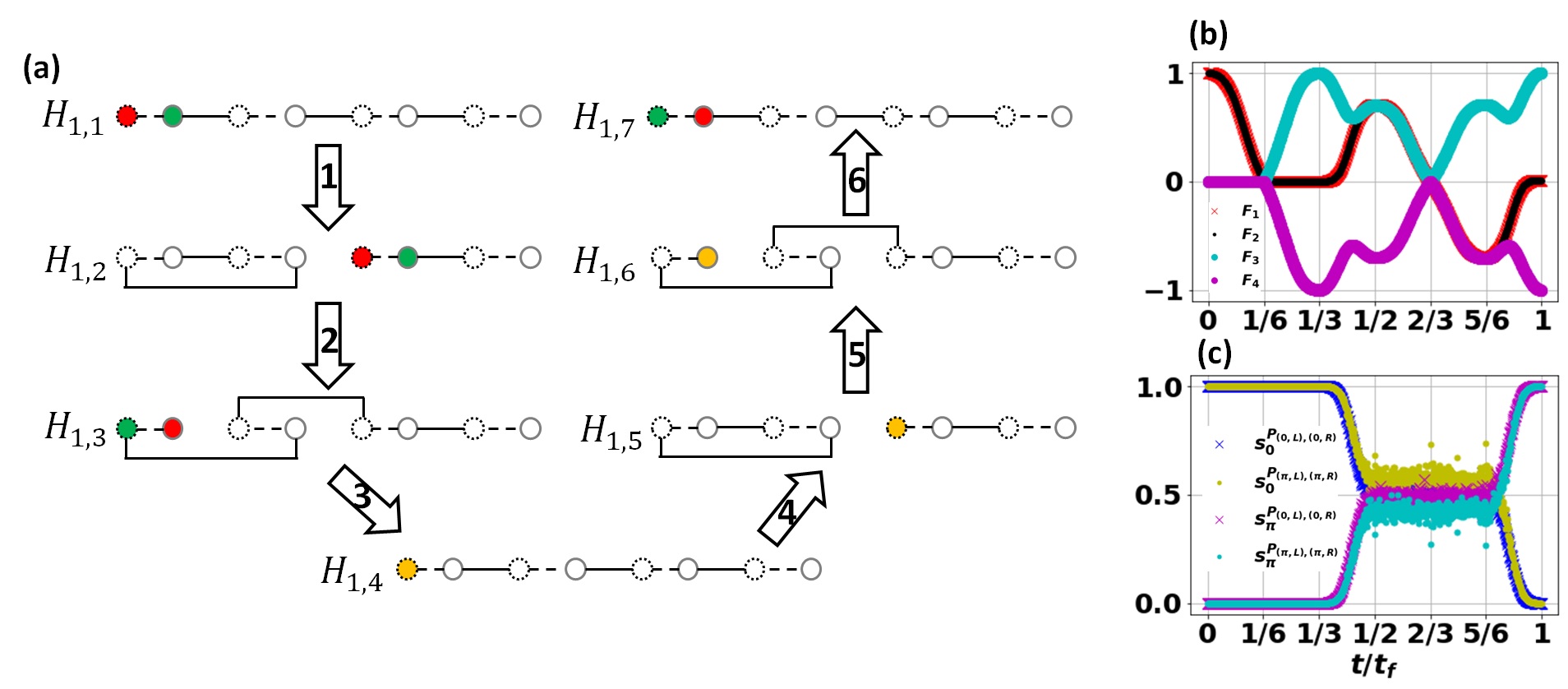}
    \caption{(a) Schematic of our six steps braiding scheme in the generalized Majorana representation of our number-conserving periodically driven $p$-wave superconductor. The generalized Majorana operators $\hat{\Gamma}_{A,j}$ and $\hat{\Gamma}_{B,j}$ are represented by dotted and solid circles respectively. The red and green filled circles represent the components of the left gMZM/gMPM that is being braided. The solid (dashed) lines represent the term appearing in $H_1$ ($H_2$) of Eq.~(\ref{Hini}). (b,c) The ``super"-fidelities $F_j(t)$ defined in the main text (panel b) and the appropriate spectral functions (panel c), plotted stroboscopically at every other period throughout our braiding protocol. In both panels, we take the truncation of the Cooper-pairs degree of freedom at $n_{\rm min} =-9$ and $n_{\rm max} =9$, $L=4$, $E_c=0$, and $\mu T = J T = \Delta T = \pi$.  }
    \label{fig:braid}
\end{figure*}    
\end{center}

It is easily verified that the left-edge localized gMZM and gMPM during the first and second steps read
\begin{eqnarray}
    \hat{\Gamma}_{0}^{(1)}(s) &=& (1-s)\hat{\Gamma}_{+,1} - s\hat{\Gamma}_{+,3} , \nonumber \\
    \hat{\Gamma}_{\pi}^{(1)}(s) &=& (1-s)\hat{\Gamma}_{-,1} - s\hat{\Gamma}_{-,3} , \nonumber \\
    \hat{\Gamma}_{0}^{(2)}(s) &=& (1-s)\hat{\Gamma}_{+,3} + s\hat{\Gamma}_{-,1} , \nonumber \\
    \hat{\Gamma}_{\pi}^{(2)}(s) &=& (1-s)\hat{\Gamma}_{-,3} - s\hat{\Gamma}_{+,1} .
\end{eqnarray}
Our braiding scheme will then turn $\hat{\Gamma}_{+,1}\rightarrow -\hat{\Gamma}_{+,3} \rightarrow -\hat{\Gamma}_{-,1}$ and $\hat{\Gamma}_{-,1}\rightarrow -\hat{\Gamma}_{-,3} \rightarrow \hat{\Gamma}_{+,1}$ by the end of step 2. Notice that while $\hat{\Gamma}_{+,1}$ and $\hat{\Gamma}_{-,1}$ are essentially exchanged by the end of step 2, the additional third step needs to be made to bring the Hamiltonian back to its original form of Eq.~(\ref{Hini}). However, we find that doing so also results in a nontrivial non-Abelian Berry phase that further maps $-\hat{\Gamma}_{-,1}\rightarrow \frac{\hat{\Gamma}_{+,1}-\hat{\Gamma}_{-,1}}{\sqrt{2}} $ and $\hat{\Gamma}_{+,1}\rightarrow \frac{\hat{\Gamma}_{+,1}+\hat{\Gamma}_{-,1}}{\sqrt{2}} $. The same three steps are thus repeated through steps 4 to 6 to yield the complete braiding, i.e., $\hat{\Gamma}_{+,1} \rightarrow -\hat{\Gamma}_{-,1} $ and $\hat{\Gamma}_{-,1}\rightarrow \hat{\Gamma}_{+,1}$ while returning the system Hamiltonian back to Eq.~(\ref{Hini}). Our braiding scheme and the evolution of its gMZM/gMPM are summarized in Fig.~\ref{fig:braid}(a).

To numerically demonstrate the above braiding scheme in our number-conserving setting, we evaluate two sets of metrics at every other period during our braiding protocol. The first set of such metrics comprises of the ``super"-fidelities
\begin{eqnarray}
F_1(t) &\equiv& \langle P_{(0,L),(0,R)} (t) | P_{(0,L),(0,R)} (0) \rangle  \rangle , \nonumber \\
F_2(t) &\equiv& \langle P_{(\pi,L),(\pi,R)} (t) | P_{(\pi,L),(\pi,R)} (0) \rangle  \rangle , \nonumber \\
F_3(t) &\equiv& \langle P_{(\pi,L),(\pi,R)} (t) | P_{(0,L),(\pi,R)} (0) \rangle  \rangle , \nonumber \\
F_4(t) &\equiv& \langle P_{(0,L),(0,R)} (t) | P_{(\pi,L),(0,R)} (0) \rangle  \rangle ,
\end{eqnarray}
where $| P_{(\alpha,L),(\beta,R)} (t) \rangle$ is a vector representation of the parity $P_{(\alpha,L),(\beta,R)} (t)$. In particular, if the braiding process is successful, it is expected that $F_1(0)=F_2(0)=1$ and $F_1(t_f)=F_2(t_f)=0$, whilst $F_3(0)=F_4(0)=0$ and $F_3(t_f)=-F_4(t_f)=1$, as indeed demonstrated in Fig.~\ref{fig:braid}(b). 

The second set of metrics we evaluate comprises of the spectral functions $s_\alpha^{P_{(A,L),(A,R)}}(t)$ with $\alpha=0,\pi$ and $A=0,\pi$, as defined in Sec.~\ref{discchar}. Intuitively, The parities $P_{(0,L),(0,R)}$ and $P_{(\pi,L),(\pi,R)}$ are both zero modes of the original Floquet operator at the beginning. If the left-edge localized gMZM and gMPM are then braided, These parities correspondingly transform into $P_{(\pi,L),(0,R)}$ and $P_{(0,L),(\pi,R)}$ respectively, which are both $\pi$ modes of the Floquet operator. Therefore, if the braiding process is successful, $s_0^{P_{(A,L),(A,R)}}(0)\approx 1$ and $s_\pi^{P_{(A,L),(A,R)}}(0)\approx 0$ for $A=0,\pi$, whilst $s_0^{P_{(A,L),(A,R)}}(t_f)\approx 0$ and $s_\pi^{P_{(A,L),(A,R)}}(t_f)\approx 1$. This is exactly what we obtained in Fig.~\ref{fig:braid}(c).

The above results show that gMZMs and gMPMs can be braided in a similar way as MZMs and MPMs in the mean-field periodically driven topological superconductors. As braiding forms a building block of topological quantum gate operations, we thus conclude that gMZMs and gMPMs found in the number-conserving periodically driven topological superconductors could also be utilized for topological quantum computing after appropriately defining their encoded qubits from the corresponding generalized Majorana parity operators.     

\section{Conclusion}
\label{conc}

In this paper, we have investigated the formation of generalized Majorana modes both in static and periodically driven $p$-wave superconductors. Similar to their mean-field counterparts, such generalized Majorana modes are topologically protected by the appropriate winding numbers, which are defined under PBCs. Indeed, we have verified that the obtained generalized Majorana modes are robust against moderate spatial disorders. Moreover, while the obtained generalized Majorana modes are no longer pinned at a specific energy/quasienergy in the presence of charging energy, we have demonstrated that the corresponding parity operators still robustly display their expected commutation relations with respect to the system's Hamiltonian (in the static case) or Floquet operator (in the time-periodic case). Consequently, such generalized Majorana modes could still yield a nonlocal qubit subspace that exhibits the same robustness as that formed by true Majorana modes. Finally, we have numerically showed that a gMZM and a gMPM emerging in our number-conserving periodically driven $p$-wave superconductor could be braided. This in turn opens up opportunities for manipulating nonlocal qubits via topologically protected quantum gates in the number-conserving topological superconductors.   

\RB{Throughout this work, we have assumed that the system's $p$-wave superconductivity arises from the proximity to an external bulk superconductor. It would be interesting to explore a different scenario in which $p$-wave superconductivity arises internally among the fermions in the semiconducting chain, either alone or on top of the above proximitized $p$-wave superconductivity. Beyond mean-field treatment, the presence of such an internal $p$-wave superconductivity amounts to adding an interacting term of the form $\sum_j \delta c_{j+1}^\dagger c_j^\dagger c_{j+1}c_j +h.c.$. The exploration of Majorana-like quasiparticles in such a system and their characterization would then require machinery from the timely field of interacting topological phases \cite{Rachel, Wu, XuMoore, Sirker_etal, Manmana_etal, Wang_etal1, Barbiero_etal, Roy_etal, Katsura_etal}, which will thus be left for future studies.}

\RB{Having uncovered the emergence of gMZMs and gMPMs in number-conserving periodically driven $p$-wave superconductors, a natural question to raise is on how they are affected by the presence of nontopological stray quasiparticles in the bulk, which may arise due to finite temperature or broadening effects. In particular, the topological and braiding properties of MZMs and MPMs in mean-field $p$-wave topological superconductors have been shown to be mostly unaffected by the presence of such quasiparticles \cite{Akhmerov10}. It is expected that the analyses presented in Ref.~\cite{Akhmerov10} could be adapted to the number-conserving case to similarly investigate the fate of the above gMZMs and gMPMs. Such an adaptation could be made, e.g., by replacing the additional coupling term between MZMs/MPMs and a stray quasiparticle of the form $\mathrm{i}\delta (a+a^\dagger)\gamma_0$ \cite{Akhmerov10} to a number conserving version $\mathrm{i}\delta (a+e^{-\mathrm{i} \hat{\phi}}a^\dagger)\Gamma_0$, the details of which will be reserved for a follow up work.}

\RB{While the presence of stray quasiparticles in the bulk will likely not significantly affect the properties of gMZMs and gMPMs, they could cause unwanted transitions outside the logical subspace encoded by the gMZMs and gMPMs and are consequently detrimental to the stored qubits. Such a quasiparticle poisoning event has remained an open problem in the area of topological quantum computing even within the mean-field framework \cite{Rainis12,march1,Lai18,Lai19,Menard19}. It leads to the finite lifetime of the Majorana-based qubits, which is found to be between $0.1\;\mu$s to $1$ ms \cite{Karzig21} for a typical mean-field topological superconductor. In this case, obtaining the lifetime estimate of the generalized Majorana-based qubits in number-conserving $p$-wave superconductors due to quasiparticle poisoning would constitute a very interesting aspect to pursue in the future.}

\RB{To mitigate the effect of quasiparticle poisoning above, many existing studies have been devoted to develop quantum error correction protocols that are compatible with (mean-field) Majorana-based qubits \cite{msur1,msur2,msur3,msur4}. In this case,} modifying and applying these existing protocols to a number-conserving setting serves as a natural and important next step. To this end, more comprehensive studies on the possibility for braiding generalized Majorana modes by means of measurements only (which are essential for executing many existing quantum error correction codes) might need to be carried out. Within the topic of topological quantum computing, it would also be fruitful to investigate the number-conserving version of interacting topological superconductors which were recently proposed to host the more exotic parafermions \cite{Bomantara2021, Mazza2018, Chew2018}. Finally, the number-conserving variations of other popular platforms that also support Majorana-modes such as second-order topological superconductors \cite{RG2020,Ghosh2021,Ghosh2021b,Vu2021,Zhou2022,Juricic2022}, non-Hermitian topological superconductors \cite{Ghosh2022,Wang2021,Liu2022,Zhou2020,Caspel2019}, and square-root topological superconductors \cite{Bomantara2022,Zhou2022b,Ezawa2020,Marques2021,Cheng2022} are expected to be worth exploring.   

 \begin{acknowledgements}
 This work was supported by the Deanship of Research
Oversight and Coordination (DROC) at King Fahd University of Petroleum \& Minerals (KFUPM) through project No.~EC221010.
 \end{acknowledgements}
	
\appendix

\section{Analytical determination of gMZMs in the number-conserving time-independent $p$-wave superconductor} 
\label{app:A}

We look for a zero energy solution to the eigenvalue equation $\mathcal{H} |E\rangle = E |E\rangle$, i.e., a solution at $E=0$, where $\mathcal{H}$ is given by Eq.~(\ref{nconsstatbdg}) in the main text. To this end, we expand $|0\rangle = \sum_j \left(A_j |\uparrow\rangle |j\rangle + B_j |\downarrow \rangle |j\rangle \right)$, where $|j\rangle$ is a basis vector representing the lattice site, and $|\uparrow/\downarrow\rangle$ is the basis vector representing the particle-hole sector. Specifically, $|\uparrow/\downarrow\rangle$ is the $+1/-1$ eigenvector of the $\sigma_z$ operator, whereas $|j\rangle$ is affected by the $\eta_x$ and $\eta_y$ operators as $\eta_x |j\rangle = |j-1\rangle + |j+1 \rangle$ and $\eta_y |j\rangle = -\mathrm{i} |j-1\rangle + \mathrm{i} |j+1 \rangle$. Moreover, we assume a semi-infinite system in which $j=1,2,\cdots, \infty$.

By explicitly evaluating $\mathcal{H} |0\rangle$ and equating the resulting coefficient of each basis state $|\uparrow/\downarrow\rangle |j\rangle$ to zero, we obtain the set of equations:
\begin{eqnarray}
    0 &=& \mu_0 A_1 + J_0 A_2 + J_0 e^{-\mathrm{i} \hat{\phi}} B_2 , \nonumber \\
    0 &=& -\mu_0 e^{-\mathrm{i} \hat{\phi}} B_1 - J_0 e^{-\mathrm{i} \hat{\phi}} B_2 - J_0 A_2 , \label{j1}
\end{eqnarray}
and 
\begin{eqnarray}
    0 &=& J_0 A_{j-1} - J_0 e^{-\mathrm{i} \hat{\phi}} B_{j-1} +\mu_0 A_j \nonumber \\
    && + J_0 A_{j+1} + J_0 e^{-\mathrm{i} \hat{\phi}} B_{j+1} , \nonumber \\
    0 &=& J_0 e^{-\mathrm{i} \hat{\phi}} B_{j-1} - J_0  A_{j-1} +\mu_0 e^{-\mathrm{i} \hat{\phi}} B_j \nonumber \\
    &&+ J_0 e^{-\mathrm{i} \hat{\phi}} B_{j+1} + J_0 A_{j+1} ,  \label{jb1}
\end{eqnarray}
for $j>1$. By adding the two lines of Eq.~(\ref{j1}), we obtain
\begin{equation}
    A_1 - e^{-\mathrm{i} \hat{\phi}} B_1 = 0 . \label{j1a}
\end{equation}
By subtracting the second line from the first line of Eq.~(\ref{jb1}), as well as adding both lines of Eq.~(\ref{jb1}) we further obtain, respectively
\begin{eqnarray}
    2J_0 (A_j - e^{-\mathrm{i} \hat{\phi}} B_j) +\mu_0 (A_{j+1} -e^{-\mathrm{i} \hat{\phi}} B_{j+1}) &=& 0, \nonumber \\
    2J_0 (A_{j+1} + e^{-\mathrm{i} \hat{\phi}} B_{j+1}) +\mu_0 (A_{j} +e^{-\mathrm{i} \hat{\phi}} B_{j}) &=& 0. \label{j1b} 
\end{eqnarray}
By plugging in Eq.~(\ref{j1a}) to the first line of Eq.~(\ref{j1b}), we obtain $(A_j - e^{-\mathrm{i} \hat{\phi}} B_j) =0$ for all $j$. On the other hand, repeatedly applying the second line of Eq.~(\ref{j1b}) gives
\begin{equation}
    A_{j+1} + e^{-\mathrm{i} \hat{\phi}} B_{j+1} = \left(-\frac{\mu_0}{2J_0} \right)^{j} (A_1 + e^{-\mathrm{i} \hat{\phi}} B_1) .
\end{equation}
It then follows that 
\begin{equation}
    |0\rangle = \sum_{j} \left(-\frac{\mu_0}{2J_0} \right)^{j} \left(|\uparrow\rangle |j\rangle + e^{\mathrm{i} \hat{\phi}}|\downarrow\rangle |j\rangle \right) .
\end{equation}
Note that such a solution is only physical and thus exists only if $|\mu_0|<|2J_0|$, under which $\left(-\frac{\mu_0}{2J_0} \right)^{j}\rightarrow 0$ as $j\rightarrow \infty$. Finally, the corresponding gMZM can be obtained by identifying $|\uparrow\rangle |j\rangle  \rightarrow c_j^\dagger $ and $|\downarrow\rangle |j\rangle \rightarrow c_j$, in which case it reads
\begin{equation}
    \hat{\Gamma}_L = \sum_{j=1}^\infty \left(-\frac{\mu_0}{2J_0} \right)^{j} \left(c_j^\dagger + e^{\mathrm{i} \hat{\phi}}c_j \right)
\end{equation}
By considering another semi-infinite system with $j=-\infty,\cdots, L$, the above procedure can be straightforwardly repeated to obtain an gMZM localized near the right end, i.e., 

\begin{equation}
    \hat{\Gamma}_R = \sum_{j=1}^\infty \mathrm{i} \left(-\frac{\mu_0}{2J_0} \right)^{j-1} \left(c_{L-j+1}^\dagger - e^{\mathrm{i} \hat{\phi}}c_{L-j+1} \right)
\end{equation}

It is worth noting that in a finite system, the gMZMs obtained from the above procedure are not exact zero energy excitations, as Eq.~(\ref{j1b}) stops at some $j$. However, as the commutators $[H,\hat{\Gamma}_L]$ and $[H,\hat{\Gamma}_R]$ can be shown to decay exponentially with $L$, $\hat{\Gamma}_L$ and $\hat{\Gamma}_R$ serve as good approximate gMZMs of the system. Finally, by recalling $\hat{\Gamma}_{A,j} \equiv e^{\mathrm{i}\hat{\phi}} c_j+c_j^\dagger$ and $\hat{\Gamma}_{B,j}\equiv \mathrm{i} \left( e^{\mathrm{i}\hat{\phi}} c_j- c_j^\dagger \right)$ defined in Sec.~\ref{ticase}, Eq.~(\ref{exactMM}) in the main text is obtained. 

\section{\RB{Commutation relation between $\hat{\phi}$ matrices and the Cooper-pairs operators}}
\label{app:exB}

\RB{In the main text, we defined two infinite dimensional matrices acting on the Cooper-pairs subspace as
\begin{eqnarray}
    \left[\Phi_x\right]_{ab} &=& \delta_{1-a,b} , \nonumber \\
    \left[\Phi_y\right]_{ab} &=& (-\mathrm{i})^{2b-1} \delta_{1-a,b} .
\end{eqnarray}
Meanwhile, by writing the Cooper-pairs operators as infinite dimensional matrices, we obtain 
\begin{eqnarray}
    \left[\cos\hat{\phi}\right]_{ab} &=& \delta_{a,b+1}+\delta_{a,b-1} , \nonumber \\
    \left[\sin\hat{\phi}\right]_{ab} &=& \mathrm{i} (\delta_{a,b+1}-\delta_{a,b-1}) .
\end{eqnarray}}

\RB{Therefore, we find that (summation over repeated indices is implied)
\begin{eqnarray}
    \left[\Phi_x \cos\hat{\phi} \Phi_x \right]_{ab} &=& [\Phi_x]_{ac} [\cos\hat{\phi}]_{cd} [\Phi_x]_{db}  \nonumber \\
    &=& \delta_{1-a,c} \left(\delta_{c,d+1}+\delta_{a,d-1}\right) \delta_{1-d,b} \nonumber \\
    &=& \delta_{1-a,2-b}+\delta_{1-a,-b} = [\cos\hat{\phi}]_{ab} , \nonumber \\
    \left[\Phi_y \cos\hat{\phi} \Phi_y \right]_{ab} &=& [\Phi_y]_{ac} [\cos\hat{\phi}]_{cd} [\Phi_y]_{db}  \nonumber \\
    &=& (-\mathrm{i})^{2(b-a)} \left(\delta_{1-a,2-b}+\delta_{1-a,-b} \right) \nonumber \\
    &=& -[\cos\hat{\phi}]_{ab} ,
\end{eqnarray}
where we have used the property of the Kronecker Delta functions that $\delta_{a,b}=\delta_{\alpha + \beta a, \alpha + \beta b}$ for any two constants $\alpha$ and $\beta$, as well as $f(a,b)\delta_{a,g(b)} = f(g(b),b)\delta_{a,g(b)}$ for any two functions $f$ and $g$. In a similar fashion, we obtain
\begin{eqnarray}
    \left[\Phi_x \sin\hat{\phi} \Phi_x \right]_{ab} &=& [\Phi_x]_{ac} [\cos\hat{\phi}]_{cd} [\Phi_x]_{db}  \nonumber \\
    &=& \mathrm{i} \delta_{1-a,c} \left(\delta_{c,d+1}-\delta_{a,d-1}\right) \delta_{1-d,b} \nonumber \\
    &=& \mathrm{i} \left(\delta_{1-a,2-b}-\delta_{1-a,-b}\right) \nonumber \\
    &=& \mathrm{i} \left(\delta_{a,b-1}-\delta_{a,b+1}\right) = -[\sin\hat{\phi}]_{ab} , \nonumber \\
    \left[\Phi_y \sin\hat{\phi} \Phi_y \right]_{ab} &=& [\Phi_y]_{ac} [\sin\hat{\phi}]_{cd} [\Phi_y]_{db}  \nonumber \\
    &=& (-\mathrm{i})^{2(b-a)+1} \left(\delta_{1-a,-b}-\delta_{1-a,2-b} \right) \nonumber \\
    &=& [\sin\hat{\phi}]_{ab} .
\end{eqnarray}}

\section{$\mathbb{Z}$ invariant in the time-independent $p$-wave superconductor} 
\label{app:B}

\RB{Starting with the usual basis in which the Pauli matrix $\sigma_z$ is diagonal, moving to a new basis in which $\mathcal{C}_1=\Phi_x\sigma_y$ is diagonal is equivalent to staying in the same basis while applying the unitary transformation $u={\rm exp}(-\mathrm{i} \frac{\pi}{4} \Phi_x \sigma_x)$ to $\mathcal{H}_k$ and all symmetry operators. Indeed, it can be easily checked that such a unitary transformation maps $\mathcal{C}_1 \rightarrow \sigma_z$, which is diagonal as intended.} On the other hand, the same unitary transformation also maps the momentum space BdG Hamiltonian to the block antidiagonal form
\begin{equation}
    \mathcal{H}_k \rightarrow \left( \begin{array}{cc}
        0 & \mathcal{W} \\
        \RB{\mathcal{W}^\dagger} & 0 
    \end{array} \right) , 
\end{equation}
where 
\begin{equation}
    \mathcal{W} = -\mathrm{i} (2J_0 \cos(k) +\mu_0) \Phi_x + 2\Delta_0 \sin(k) e^{-i\hat{\phi}} .  
\end{equation}
We may now evaluate the winding number 
\begin{equation}
    W = \frac{1}{2\pi \mathrm{i}} \oint {\rm Tr}\left(\mathcal{W}^{-1} d\mathcal{W} \right)  ,
\end{equation}
where the trace operates on the Hilbert space of Cooper-pairs. We find that 
\begin{equation}
    {\rm Tr}\left(\mathcal{W}^{-1} d\mathcal{W} \right) = \frac{1}{2}\left\lbrace \frac{1}{z_+ +\mu_0} dz_+ + \frac{1}{z_-+\mu_0} dz_- \right\rbrace {\rm Tr}\left(\mathcal{I}_c\right) , \label{traceex} 
\end{equation}
where $z_\pm =2J_0 \cos(k) \pm \mathrm{i} 2 \Delta_0 \sin(k)$ and $\mathcal{I}_c$ is the identity in the Cooper-pairs subspace. By recalling that the normalized winding number is defined as 
\begin{equation}
    w = \frac{W}{{\rm Tr}\left(\mathcal{I}_c\right)} ,
\end{equation}
together with Eq.~(\ref{traceex}), Eq.~(\ref{statwinding}) in the main text is obtained.

\section{$\mathbb{Z}\times \mathbb{Z}$ invariant definition in the time-periodic $p$-wave superconductor} 
\label{app:C}

We first note that Eq.~(\ref{usym}) in the main text can be written in the form
\begin{equation}
    \mathcal{U}_{k,\rm sym} = F G ,
\end{equation}
where 
\begin{eqnarray}
    F&=& e^{-\mathrm{i} \mu \sigma_z \frac{T}{4}} e^{-\mathrm{i} \left[ 2J \cos(k) \sigma_z - 2\Delta \sin(k)  \left(e^{-\mathrm{i} \hat{\phi}} \sigma_+ - e^{\mathrm{i} \hat{\phi}} \sigma_- \right) \right] \frac{T}{4} } ,\nonumber \\
    G &=& \mathcal{C}_1 F^\dagger \mathcal{C}_1^\dagger = \mathcal{C}_2 F^\dagger \mathcal{C}_2^\dagger \nonumber \\
    &=& e^{-\mathrm{i} \left[ 2J \cos(k) \sigma_z - 2\Delta \sin(k)  \left(e^{-\mathrm{i} \hat{\phi}} \sigma_+ - e^{\mathrm{i} \hat{\phi}} \sigma_- \right) \right] \frac{T}{4} } e^{-\mathrm{i} \mu \sigma_z \frac{T}{4}} \;. \nonumber \\
\end{eqnarray}
In the canonical basis where $\mathcal{C}_1 = {\rm diag}(\mathcal{I}_c, -\mathcal{I}_c)$, the matrix representation for $F$ can be written as
\begin{equation}
    F \hat{=} \left(\begin{array}{cc}
       A  &  B\\
       C  &  D
    \end{array}\right),
\end{equation}
where $A$, $B$, $C$, and $D$ are infinite dimensional matrices acting on the Cooper-pairs subspace. In particular, at $J=\Delta$,
\begin{eqnarray}
    B &=& - \left[\cos(k)\cos\left(\tilde{\mu}\right) \sin\left(\tilde{J} \right) + \cos\left(\tilde{J} \right) \sin\left(\tilde{\mu} \right) \right] \Phi_x \nonumber \\
    &&  +\mathrm{i} \sin(k) \cos\left(\tilde{\mu}\right) \sin \left(\tilde{J} \right) e^{-\mathrm{i} \hat{\phi}} \;, \\
    D &=& -\sin\left(\tilde{\mu}\right) \sin\left(\tilde{J} \right) (\cos(k) \mathcal{I}_c -\mathrm{i} \sin(k) \Phi_x e^{-\mathrm{i} \hat{\phi}})  \nonumber \\
    && +\cos\left(\tilde{\mu}\right) \cos\left(\tilde{J} \right) \mathcal{I}_c, 
\end{eqnarray}
where $\tilde{\mu}=\mu T/4$ and $\tilde{J}=J T/2$. Generalizing the results of Ref.~\cite{Asboth2014,RG2020b}, we define the normalized winding numbers 
\begin{eqnarray}
    w_0 &=& \frac{1}{2\pi \mathrm{i} {\rm Tr}(\mathcal{I}_c)} \oint \mathrm{Tr}\left(B^{-1} dB\right)   \nonumber \\
    &=& \sum_{s=\pm}\frac{1}{4\pi \mathrm{i}} \oint \frac{1}{z_{s,0} +\cos(JT/2)\sin(\mu T/4)} dz_{s,0} \;, \nonumber \\
    && \\
    w_\pi &=& \frac{1}{2\pi \mathrm{i} {\rm Tr}(\mathcal{I}_c)} \oint \mathrm{Tr}\left(D^{-1} dD\right)  \nonumber \\
    &=& \sum_{s=\pm}\frac{1}{4\pi \mathrm{i}} \oint \frac{1}{z_{s,\pi} +\cos(JT/2)\cos(\mu T/4)} dz_{s,\pi} \;. \nonumber \\
\end{eqnarray}
where $z_{\pm,0} = e^{\pm \mathrm{i} k} \cos(\mu T/4)\sin(JT/2)$ and $z_{\pm,\pi} = e^{\pm \mathrm{i} k} \sin(\mu T/4)\sin(JT/2)$. In this case, $w_0$ ($w_\pi$) counts the number of pairs of edge-localized zero ($\pi/T$) quasienergies of $\mathcal{U}_{k,\rm sym}$, which thus translates to the number of pairs of gMZMs (gMPMs).   

\section{Proof of Eq.~(\ref{idgam}) in the main text}
\label{app:D}

We first note that $[\hat{\Gamma}_{A,j+1}^\dagger \hat{\Gamma}_{B,j},\hat{\Gamma}_{A,j'+1}^\dagger \hat{\Gamma}_{B,j'}] = 0 $. Therefore,
\begin{equation}
    e^{\sum_j  \theta \hat{\Gamma}_{A,j+1}^\dagger \hat{\Gamma}_{B,j} } = \prod_j e^{ \theta \hat{\Gamma}_{A,j+1}^\dagger \hat{\Gamma}_{B,j} } . \label{appdchk1}
\end{equation}
Next, as $\left(\mathrm{i} \hat{\Gamma}_{A,j+1}^\dagger \hat{\Gamma}_{B,j}\right)^2=1$, Taylor expansion gives an Euler-like formula for
\begin{equation}
    e^{ \theta \hat{\Gamma}_{A,j+1}^\dagger \hat{\Gamma}_{B,j} } = \cos(\theta) +\sin(\theta) \hat{\Gamma}_{A,j+1}^\dagger \hat{\Gamma}_{B,j} .
\end{equation}
Consequently,
\begin{eqnarray}
    e^{ \theta \hat{\Gamma}_{A,j+1}^\dagger \hat{\Gamma}_{B,j} } \hat{\Gamma}_{A,j'} e^{- \theta \hat{\Gamma}_{A,j+1}^\dagger \hat{\Gamma}_{B,j} } &=& \nonumber \\ \begin{cases}
    \hat{\Gamma}_{A,j'} & {\rm \;\;for\;\;} j'\neq j+1 \\
    \cos(2\theta) \hat{\Gamma}_{A,j+1} -\sin(2\theta) \hat{\Gamma}_{B,j} & {\rm \;\;for\;\;} j'= j+1
    \end{cases} && \nonumber \\
    e^{ \theta \hat{\Gamma}_{A,j+1}^\dagger \hat{\Gamma}_{B,j} } \hat{\Gamma}_{B,j'}^\dagger e^{- \theta \hat{\Gamma}_{A,j+1}^\dagger \hat{\Gamma}_{B,j} } &=& \nonumber \\ \begin{cases}
    \hat{\Gamma}_{B,j'}^\dagger & {\rm \;\;for\;\;} j'\neq j \\
    \cos(2\theta) \hat{\Gamma}_{B,j}^\dagger +\sin(2\theta) \hat{\Gamma}_{A,j+1}^\dagger & {\rm \;\;for\;\;} j'= j
    \end{cases} && \label{appdchk2}
\end{eqnarray}
Finally, Eqns.~(\ref{appdchk1}) and (\ref{appdchk2}) give
\begin{eqnarray}
    e^{ \sum_j \theta \hat{\Gamma}_{A,j+1}^\dagger \hat{\Gamma}_{B,j} } \hat{\Gamma}_{A,\ell+1} e^{- \sum_j \theta \hat{\Gamma}_{A,j+1}^\dagger \hat{\Gamma}_{B,j} } &=& \nonumber \\ 
    \left(\prod_j e^{ \theta \hat{\Gamma}_{A,j+1}^\dagger \hat{\Gamma}_{B,j} } \right) \hat{\Gamma}_{A,\ell+1} \left( \prod_j e^{- \theta \hat{\Gamma}_{A,j+1}^\dagger \hat{\Gamma}_{B,j} } \right) &=& \nonumber \\
    e^{ \theta \hat{\Gamma}_{A,\ell+1}^\dagger \hat{\Gamma}_{B,\ell} } \hat{\Gamma}_{A,\ell+1}  e^{- \theta \hat{\Gamma}_{A,\ell+1}^\dagger \hat{\Gamma}_{B,\ell} } &=& \nonumber \\
    \cos(2\theta) \hat{\Gamma}_{A,\ell+1} -\sin(2\theta) \hat{\Gamma}_{B,\ell} && 
\end{eqnarray}
and
\begin{eqnarray}
    e^{ \sum_j \theta \hat{\Gamma}_{A,j+1}^\dagger \hat{\Gamma}_{B,j} } \hat{\Gamma}_{B,\ell}^\dagger e^{- \sum_j \theta \hat{\Gamma}_{A,j+1}^\dagger \hat{\Gamma}_{B,j} } &=& \nonumber \\ 
    \left(\prod_j e^{ \theta \hat{\Gamma}_{A,j+1}^\dagger \hat{\Gamma}_{B,j} } \right) \hat{\Gamma}_{B,\ell}^\dagger \left( \prod_j e^{- \theta \hat{\Gamma}_{A,j+1}^\dagger \hat{\Gamma}_{B,j} } \right) &=& \nonumber \\
    e^{ \theta \hat{\Gamma}_{A,\ell+1}^\dagger \hat{\Gamma}_{B,\ell} } \hat{\Gamma}_{B,\ell}^\dagger  e^{- \theta \hat{\Gamma}_{A,\ell+1}^\dagger \hat{\Gamma}_{B,\ell} } &=& \nonumber \\
    \cos(2\theta) \hat{\Gamma}_{B,\ell}^\dagger +\sin(2\theta) \hat{\Gamma}_{A,\ell+1}^\dagger &,& 
\end{eqnarray}
thus recovering Eq.~(\ref{idgam}) in the main text.

\end{document}